\documentclass[format=acmsmall, review=false, screen=true]{acmart}

\usepackage{fancyhdr}
\usepackage{subfig}
\usepackage{graphicx}
\usepackage{array}
\usepackage{url}
\usepackage{footmisc} 
\usepackage{fancybox}
\usepackage{multirow}
\usepackage{amssymb}
\usepackage{color}
\usepackage{colortbl}
\usepackage{subfig}
\usepackage{diagbox}
\usepackage{bbding}
\usepackage{placeins}
\usepackage{balance}
\usepackage{listings}
\usepackage{amsmath}
\usepackage{amsfonts}
\usepackage{stackrel}
\usepackage{svg}
\usepackage{setspace}
\usepackage{natbib}
\setcitestyle{square,comma,numbers}
\PassOptionsToPackage{prologue,dvipsnames}{xcolor}
\usepackage[dvipsnames]{xcolor}
\usepackage[most]{tcolorbox}
\usepackage[linesnumbered, ruled,vlined]{algorithm2e}
\usepackage{hyperref}
\newcommand{\ie}{\textit{i.e.,} }

\usepackage{amssymb}
\usepackage{makecell}

\usepackage[utf8]{inputenc}
\usepackage{tcolorbox}
\tcbuselibrary{breakable}
\usepackage{geometry}
\usepackage{wrapfig}
\usepackage{tikz}
\newcommand{\blackcircled}[1]{%
    \tikz[baseline=(char.base)]{%
        \node[shape=circle,draw=black,fill=black,inner sep=1pt,text=white] (char) {\small#1};%
    }%
}

\newcommand{\todo}[1]{}
\renewcommand{\todo}[1]{{\color{red} TODO: {#1}}}

\begin{document}
       
	\title{An Empirical Study of Agent Developer Practices in AI Agent Frameworks}

	\author{Yanlin Wang}
	\affiliation{%
		\institution{Sun Yat-sen University}
            \country{China}
	}
	\email{wangylin36@mail.sysu.edu.cn}
    
         \author{Xinyi Xu}
	\affiliation{%
		\institution{Sun Yat-sen University}
            \country{China}
	}
        \email{xuxy299@mail2.sysu.edu.cn}
        
        \author{Jiachi Chen}
	\affiliation{%
		\institution{Zhejiang University}
        \country{China}
	}
	\email{chenjch86@mail.sysu.edu.cn}
	
	\author{Tingting Bi}
	\affiliation{%
		\institution{The University of  Melbourne}
            \country{Australia}
	}
	\email{Tingting.Bi@unimelb.edu.au}

\author{Wenchao Gu}
	\affiliation{%
		\institution{Technical University of Munich}
            \country{Germany}
	}
	\email{wenchao.gu@tum.de}

        \author{Zibin Zheng}
	\affiliation{%
		\institution{Sun Yat-sen University}
            \country{China}
	}
	\email{zhzibin@mail.sysu.edu.cn}

	\UseRawInputEncoding
\pdfoutput=1
\begin{abstract}
The rise of large language models (LLMs) has sparked a surge of interest in agents, leading to the rapid growth of agent frameworks. Agent frameworks are software toolkits and libraries that provide standardized components, abstractions, and orchestration mechanisms to simplify agent development. More than 100 open-source agent frameworks have emerged on GitHub, collectively accumulating over 400,000 stars and 70,000 forks. Despite widespread use of agent frameworks, their practical applications and how they influence the agent development process remain underexplored. Different agent frameworks encounter similar problems during use, indicating that these recurring issues deserve greater attention and call for further improvements in agent framework design. Meanwhile, as the number of agent frameworks continues to grow and evolve, more than 80\% of developers report difficulties in identifying the frameworks that best meet their specific development requirements.

In this paper, we conduct the first empirical study of LLM-based agent frameworks, exploring real-world experiences of developers in building AI agents. Specifically, we collect and analyze 1,575 LLM-based agent projects on GitHub along with 8,710 related developer discussions. Based on this, we identify ten representative frameworks and examine the functions they fulfill in the development process, how developers use them, and their popularity trends. In addition, we construct a taxonomy of agent development challenges across the software development lifecycle, 
covering four categories and nine distinct subcategories.

To compare how well the agent frameworks meet developer needs, we further collect developer discussions for the ten previously identified agent frameworks, 
resulting in a total of 11,910 discussions. 
Finally, by analyzing these discussions, we compare the frameworks across five dimensions: development efficiency, which reflects how effectively a framework accelerates coding, debugging, and prototyping; functional abstraction, which concerns the clarity and modularity of the framework design in simplifying complex agent behaviors; learning cost, which captures the difficulty developers face in acquiring the knowledge needed to use the framework; performance optimization, which describes how well the framework manages computational resources during execution; and maintainability, which refers to how easily developers can update and extend both the framework itself and the agents built upon it over time.
Our comparative analysis reveals significant differences among frameworks in how they meet the needs of agent developers.
Overall, we provide a set of findings and implications for the LLM-driven AI agent framework ecosystem and offer insights for the design of future LLM-based agent frameworks and agent developers. 
\end{abstract}

	

\maketitle
\UseRawInputEncoding
\pdfoutput=1
\section{Introduction}
\label{Introduction}

Agent frameworks are software toolkits and libraries that provide standardized components, abstractions, and orchestration mechanisms to simplify the construction of agents ~\cite{martin1999open}. 
With the rapid progress of large language models (LLMs), the landscape of agent development has expanded substantially. More than 100 open-source agent frameworks have emerged on GitHub \cite{vinlam2025agents}, collectively accumulating over 400,000 stars and 70,000 forks. These frameworks enable developers to build autonomous agents for diverse purposes such as reasoning, tool use, and collaboration \cite{ wang2024agent, rasheed2024autonomous,dou2025pre}.

Research on intelligent agents has made significant progress, primarily focusing on architectural design and application domains of agents, system robustness, and multi-agent coordination \cite{joshi2025comprehensive,cemri2025multi,lu2024llms,kolt2025governing}. However, how agent development frameworks affect agent developers and the overall agent development process is still underexplored. Agent development, like conventional software engineering, follows the software development lifecycle (SDLC), which spans stages such as design, implementation, testing, deployment, and maintenance.
The lack of understanding of how agent development frameworks influence the agent development process has led developers to face various challenges at different stages of the SDLC.

To fill this gap, we conduct the first empirical study to understand the challenges that developers encounter when building agents using frameworks. 
Our empirical study seeks to inform the optimization directions for agent framework designers, and further supports agent developers in selecting frameworks that are best suited to their specific requirements.
To this end, we collect 1,575 LLM-based agent projects and 8,710 related developer discussions, from which we identify ten widely used agent frameworks. We then further collect and analyze 11,910 developer discussions specifically related to these frameworks.

Through this empirical investigation, we aim to answer the following three research questions that guide our study.

\textbf{RQ1. How are LLM-based agent frameworks adopted and utilized by developers in real-world projects?}

To lay a foundational understanding of the current agent framework landscape, we first aim to uncover how LLM-based agent development frameworks are actually adopted and utilized by developers in practice. Specifically, we analyze the specific roles they play in real-world applications, their adoption across projects, and trends in popularity over time.
Specifically, we examine 8,710 discussion threads from three perspectives: \textbf{(i)} functional roles, meaning the kinds of development tasks each framework primarily supports; \textbf{(ii)} usage patterns, which describe the typical ways frameworks are integrated in practice; and \textbf{(iii)} community popularity, refers to the level of developer discussion and attention.
Our findings reveal that when selecting an agent framework, developers should prioritize ecosystem robustness, long-term maintenance activity, and practical adoption in real projects, rather than focusing solely on short-term community popularity. 
We also observe that combining multiple frameworks with different functions has become the dominant approach to agent development.

\textbf{RQ2. What challenges do developers face across the SDLC when building AI agents with frameworks?}

Building on the foundational landscape of frameworks and usage trends established in RQ1, to further understand the challenges when developers use these frameworks in real-world development, we extract challenges from 8,710 GitHub discussions and classify them into stages of the software development lifecycle (SDLC). Our results show that developers encounter challenges in four main domain:
\textbf{(i)} Logic: Logic-related failures summarize those faults that originate from deficiencies in the agent's
internal logic control mechanisms. This domain primarily aligns with the stages of the software development lifecycle, from early design to deployment. Issues related to task termination policies account for 25.6\%, reflecting insufficient support in current frameworks for task flow management and infinite loop prevention.
\textbf{(ii)} Tool: Tool-related failures summarize those faults that originate from deficiencies in the agent's tool integration and interaction mechanisms.
Spanning the implementation, testing, and deployment stages, tool-related issues account for 14\% of reported problems. Examples include API limitations, permission errors, and third-party library mismatches.
\textbf{(iii)} Performance: 
Performance-related failures summarize those faults that originate from inefficiencies or limitations in the agent's resource management and execution performance.
Emerging during implementation to deployment phases, these challenges represent 25\% of all challenges and manifest as failures in context retention, loss of dialogue history, and memory management errors. 
\textbf{(iv)} Version: 
Version-related failures summarize those faults that originate from incompatibilities caused by agent framework or dependency version updates.
Corresponding to the deployment, testing, and maintenance stage of the SDLC, this domain affects over 23\% of projects. Version conflicts and compatibility issues often cause build failures, thereby increasing development and maintenance costs.

\textbf{RQ3. How do different frameworks differ in meeting developers' needs?}

Following the identification of SDLC challenges in RQ2, we now focus on evaluating how effectively existing agent development frameworks address these practical challenges and satisfy developers' needs. To do this, we analyze the framework's support from five aspects: learning cost assesses how difficult it is for developers to acquire the knowledge needed to build and deploy AI agent; development efficiency evaluates how effectively a framework accelerates coding, debugging, and prototyping by reducing redundant or repetitive work; functional abstraction explores how well the framework transforms complex technical steps into easy-to-use and reusable building blocks; performance optimization reflects the framework's capability to utilize computational resources efficiently during execution; maintainability captures how easily agent developers can continuously update and enhance both the framework itself and the applications built upon it over time. Specifically, we find that \textbf{(i)} Langchain and CrewAI lower the technical threshold for beginners. \textbf{(ii)} AutoGen and LangChain excel at rapid prototyping. \textbf{(iii)} In terms of functional encapsulation, AutoGen and LangChain are leading in task decomposition and multi-agent collaboration. \textbf{(iv)} Performance optimization is a common shortcoming across
all frameworks. \textbf{(v)} Despite their mature ecosystems, AutoGen and LangChain
face the highest maintenance complexity.

We summarize our main contributions as follows: 
\begin{itemize} 
\item The first large-scale empirical investigation and analysis of ten widely used agent frameworks highlights the challenges, real-world applications, and functional roles of these frameworks in the agent development process.
\item A taxonomy of development challenges in the software development lifecycle (SDLC), covering four domain and nine categories.
\item Development of a five-dimensional evaluation metric based on agent developer needs is carried out to compare the ten frameworks.
\item Summary of key findings and practical suggestions for agent stakeholders, including agent developers and agent framework designers.
\end{itemize}
\textbf{Paper Organization.} The rest of this paper is organized as follow. Section~\ref{sec:background} introduces the background of the agent framework and software development life cycle. 
Section~\ref{sec:data} presents the data collection and processing procedures for RQ1, RQ2, and RQ3.
Section~\ref{sec:RQ1} presents the results for RQ1, focusing on framework popularity and usage trends. Section~\ref{sec:RQ2} analyze the challenges encountered during agent development. In Section~\ref{sec:RQ3}, we provide a detailed comparative analysis of the 10 frameworks to answer RQ3. Finally, Section~\ref{sec:discussion} discusses the implications of our findings, Section~\ref{sec:related} reviews related work on agent frameworks and empirical software engineering studies, and Section~\ref{sec:conclusion} concludes the paper and outlines future work.

\UseRawInputEncoding
\pdfoutput=1
\section{Background}
\label{sec:background}
This section provides a brief background on LLM-based agents, agent frameworks, and the software development life cycle.

\subsection{LLM-Based Agent}
An LLM-based agent is an artificial intelligence system built around a Large Language Model (LLM) as its core~\cite{wang2025agents}, capable of autonomously performing complex task. Large Language Models~\cite{chang2024survey,patil2024gorilla,dou2025evalearn} are deep neural networks trained on massive text corpora to learn world knowledge through self-supervised learning~\cite{zhao2024comparison}. They exhibit remarkable capabilities in code translation~\cite{codetrans1,codetrans2,codetrans3,codetrans4}, code search~\cite{codesearch1,codesearch2,hu2024tackling,codesearch4,codesearch5,codesearch6}, code summarization~\cite{codesum1,codesum2,shi-etal-2021-cast,codesum4}, vulnerability detection~\cite{vulner1,vulner2,vulner3,vulner4,vulner5,vulner6}, issue resolution~\cite{issue1,issue2,issue3} and generation ~\cite{codegen1,codegen2,codegen3,codegen4,codegen5,codegen6,codegen7,codegen8,codegen9,codegen10,codegen11,codegen12,codegen13,codegen14,codegen15,codegen16,codegen17}. Unlike standard LLMs that generate a single response to a given prompt~\cite{schulhoff2024prompt}, LLM-based agents can comprehend intricate contexts~\cite{lee2024humaninspiredreadingagentgist}, plan sequential actions~\cite{yao2023react}, and interact dynamically with humans and external environments~\cite{maggioni2021learninginteractionkernelsagent}. Typically, an LLM-based agent consists of four essential components: the brain, memory~\cite{zhong2024memorybank}, planning~\cite{valmeekam2023planbench}, and tools~\cite{patil2024gorilla,cai2024latm,lewis2020rag}. The brain refers to the LLM itself, which is responsible for language understanding and generation~\cite{cemri2025multi} . The memory component enables the agent to enhance task continuity by recalling and utilizing past experiences. The planning mechanism allows the agent to decompose complex objectives into executable subgoals and formulate actionable strategies~\cite{qiao2024agent}. The tools component enables the agent to interact with external systems, thereby performing operations that extend beyond text generation~\cite{masterman2024landscape}.
In the early stages of development, LLM-based agents are primarily implemented through manual construction. Developers often design the agent's architecture and logic from scratch. Although this manual development approach offer flexibility, Beginners may find it hard to quickly build agents using it.
To address these limitations and enhance the efficiency of agent development and deployment, researchers gradually introduce the agent frameworks. 

\subsection{Agent Frameworks}
LLM-based frameworks are toolkits or platforms designed for constructing, managing, and coordinating intelligent agents~\cite{huang2025ai}. These frameworks manage key aspects of agents such as planning, decision-making, and memory, while enabling interaction with other systems and tools to accomplish complex tasks~\cite{hasan2025empiricalstudytestingpractices}. In general, such frameworks simplify the development process allowing researchers to focus on high-level task reasoning~\cite{dou2025improving} and strategy design without the need to repeatedly implement foundational ~\cite{yao2023react,cai2024latm,patil2024gorilla}components. As of October 2025, more than one hundred agent frameworks have been proposed on open-source platforms for developers to explore and utilize~\cite{agentproject_exhaustive_list,vinlam_agents_table}.
In this section, we primarily introduce the ten agent frameworks identified and examined in this study.

\begin{table*}[htbp]
\centering
\footnotesize
\setlength{\tabcolsep}{4pt}
\caption{The Ten LLM-based Agent Frameworks: Functions and Core Mechanisms}
\begin{tabular}{@{}p{1.4cm}|p{6.6cm}|p{6.8cm}@{}}
\toprule
\textbf{Framework} & \textbf{Function} & \textbf{Mechanism} \\
\midrule
LangChain & Simplifies the development of LLM-based agents. & Orchestrates components using sequential and hierarchical chain structures. \\
LangGraph & Manages complex agent workflows. & Represents reasoning and execution paths as a Directed Acyclic Graph (DAG). \\
AutoGen & Enables multi-agent collaboration for complex problem-solving. & Implements a ``conversation-as-computation'' paradigm through inter-agent dialogues. \\
CrewAI & Organizes cooperative multi-agent workflows for task automation. & Utilizes a hierarchical crew structure with predefined roles and task assignments. \\
MetaGPT & Automates end-to-end project generation and management. & Encodes human Standard Operating Procedures (SOPs) into agent role interactions. \\
LlamaIndex & Augments LLMs with access to external knowledge sources. & Employs Retrieval-Augmented Generation (RAG) to ground responses in retrieved data. \\
Swarm & Facilitates large-scale, decentralized multi-agent collaboration. & Abstracts multi-agent coordination through a centralized controller and message-passing orchestration layer. \\
BabyAGI & Performs autonomous task planning and iterative execution. & Operates a self-improving task loop for generation, prioritization, and evaluation. \\
Camel & Explores emergent cooperation in large multi-agent populations. & Implements role-playing communication between agents to simulate cooperation and analyze emergent behaviors. \\
Semantic Kernel & Integrates agent capabilities into enterprise and software systems. & Leverages a modular, plugin-based architecture for extensibility and interoperability \\
\bottomrule
\end{tabular}
\label{tab:agent_frameworks}
\end{table*}

LangChain ~\cite{mavroudis2024langchain} is one of the earliest  orchestration frameworks, which lowers the barrier for developers building agents by providing a range of modular components such as Models and Chains ~\cite{annam2025langchain}. However, its initial ``chain-based'' design struggles to handle scenarios requiring state management, such as loops, which directly leads to the development of LangGraph. LangGraph ~\cite{wang2024agent}makes it possible to build workflows with complex logic by modeling the agent workflow as a directed acyclic graph . AutoGen is a framework that enables collaborative task completion through multi-agent conversations. It introduces the concept of ``conversation as computation,'' modeling tasks as dialogues among agents ~\cite{wu2024autogen}. 
CrewAI ~\cite{moura2024crewai} focuses on executing structured workflows by assigning ``roles'' and ``tasks'' to agents and organizing them into a ``crew''. MetaGPT compares the entire task-solving process to a software company's project development ~\cite{hong2024metagpt}. By simulating human SOPs, it generates more complete solutions than chat-based multi-agent systems. LlamaIndex focuses on enhancing agent with retrieval-augmented generation (RAG), which enables them to access, retrieve, and incorporate external knowledge sources for more factual and context-grounded responses \cite{lewis2020rag,llamaindex2024doc}. In addition, several experimental or domain-specific frameworks have emerged: Swarm explores multi-agent interfaces~\cite{swarm2024github}, BabyAGI demonstrates autonomous task planning and execution loops based on a single agent~\cite{nakajima2023babyagi}, Camel investigates scaling laws of multi-agent systems~\cite{li2023camel}, and Semantic Kernel emphasizes deep integration with enterprise ecosystems such as Python and Java, highlighting maintainability and extensibility~\cite{microsoft2024semantickernel}.

\subsection{Software Development Life Cycle}
The Software Development Life Cycle (SDLC) ~\cite{ruparelia2010software} provides a systematic process for planning, designing, developing, testing, and maintaining software applications. In the era of large language models (LLMs), agents are regarded as a new form of software, and their development also follows a structured life-cycle model ~\cite{rasheed2024autonomous}. Thus, the development of agents can be naturally mapped onto the stages of the traditional SDLC.
The process begins with requirements analysis, which defines the agent's functions, performance, and constraints. This stage also establishes clear goals and scope. The design phase specifies the agent's architecture, module composition, interaction flows, and data models. The development phase transforms the design into a functional agent system, involving coding, model integration, and tool implementation. The testing phase ensures that the agent system meets functional, performance, and security requirements through unit, integration, system, and acceptance testing, with special consideration of the agent's autonomy and potential unpredictability. The deployment phase delivers the tested agent system into a production environment to achieve stable, secure, and reliable operation. Finally, the maintenance phase provides continuous support after deployment, addressing issues, applying updates, and adding new features to ensure that the agent remains functional over time.

\UseRawInputEncoding
\pdfoutput=1
\section{Data Collection and Processing}
\label{sec:data}
To capture developer feedback and understand the real-world usage of agent framework, we construct a dataset of 1,575 GitHub repositories related to agents to support our study. 
\begin{figure*}[t]
    \centering
\includegraphics[width=0.9\linewidth, keepaspectratio]{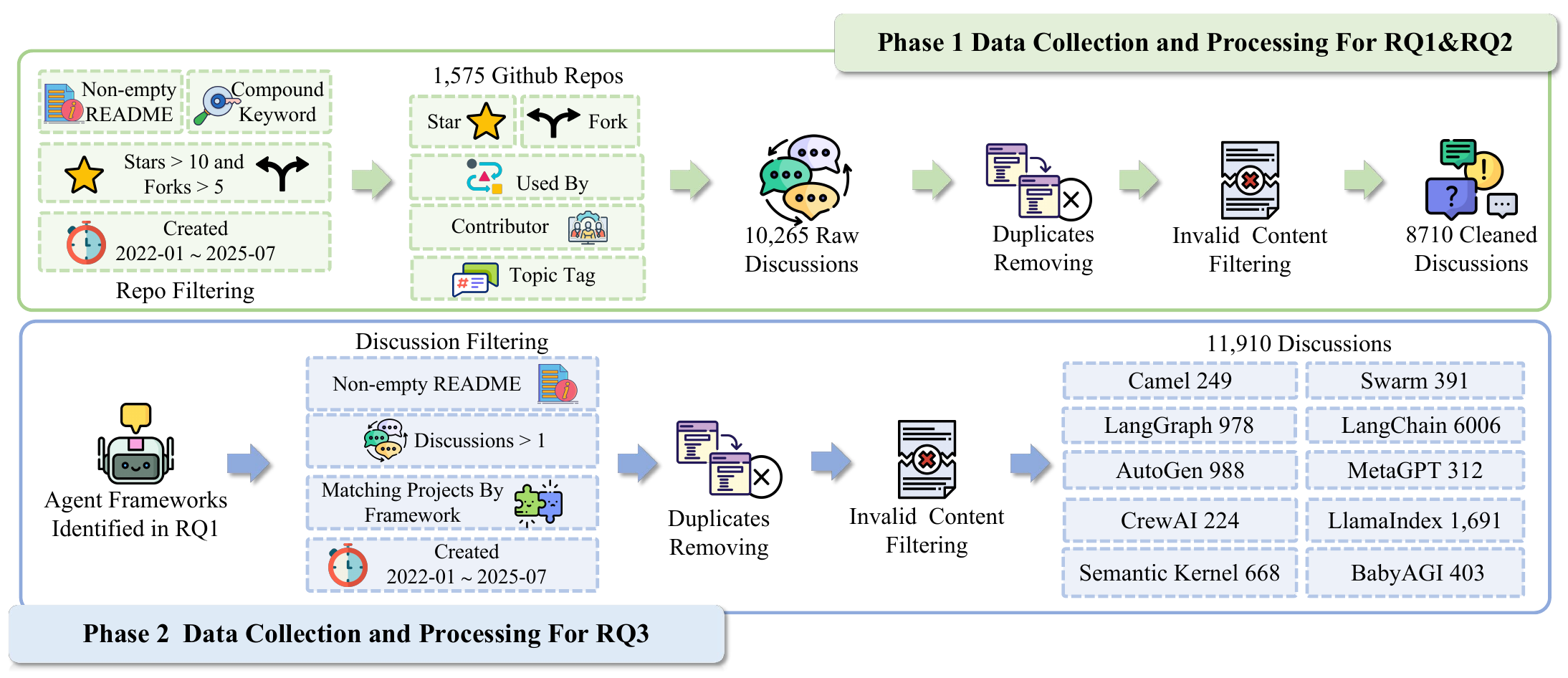}
    \vspace{-0.5em}
    \caption{Pipeline for Data Collection and Processing in RQ1-RQ3.}
    \vspace{-1em}
    \label{fig:framework_trends}
\end{figure*}
The repository selection process follows the criteria below to ensure relevance and quality:
\textbf{(i) Keyword inclusion.} To balance coverage and relevance, we adopt a compound keyword filtering strategy. The project name or description must include the keywords:
\texttt{``agent'' AND (``AI'' OR ``LLM'' OR ``assistant'')}.
This approach effectively reduces unrelated repositories such as system or monitoring agents and ensures that the selection focuses on LLM-based agent frameworks.
\textbf{(ii) Documentation completeness.} Each project must provide a non-empty \texttt{README} file to ensure basic documentation quality.
\textbf{(iii) Community attention.} Each project must have more than 10 stars and 5 project copies (forks) to indicate a minimal level of community engagement.
\textbf{(iv) Recency.} Only repositories created or updated between 2022 and July 2025 are included to ensure technical relevance.
For each project, we collect metadata including the number of stars, project copies (forks), dependent repositories (used by), contributors, topic labels, and discussion threads. In total, we obtain \textbf{10,265} discussion threads. Each record contains the topic title and body, all replies and follow-up comments, timestamps, author roles, and associated project metadata.

In addition, to ensure data quality and analytical validity, we apply cleaning process as follows:
\textbf{(i) Remove duplicates.} Discussions with identical titles or content, including cross-posted or mirrored topics across repositories, are removed.
\textbf{(ii) Filter invalid content.} We exclude empty messages, emoji-only posts, purely promotional or non-informative comments (\ie ``Great project!''), and bug reports lacking sufficient context for meaningful interpretation.
After cleaning, we retain \textbf{8,710} discussions, which serve as the foundation for subsequent analyses in RQ1 and RQ2.

Futhermore, to compare how well the ten agent frameworks identified in RQ1 meet developer needs, we further collect all discussion threads from projects built upon these ten frameworks. The project selection process follows the criteria below:
\textbf{(i) Keyword inclusion.} The project name or description must include the framework name (\ie ``LangChain'').
\textbf{(ii) Documentation completeness.} Each project must provide a non-empty \texttt{README} file to ensure proper documentation.
\textbf{(iii) Discussion activity.} Each project must contain at least one discussion thread.
\textbf{(iv) Recency.} Only repositories created or updated between 2022 and July 2025 are included to ensure technical relevance.
Following the same cleaning process, we finally obtain a total of \textbf{11,910} discussions, distributed as 6,006 for LangChain, 978 for LangGraph, 988 for AutoGen, 224 for CrewAI, 391 for Swarm, 249 for Camel, 312 for MetaGPT, 1,691 for LlamaIndex, 668 for Semantic Kernel, and 403 for BabyAGI.

\begin{figure*}[t]
    \centering
    \includegraphics[width=0.9\linewidth, keepaspectratio]{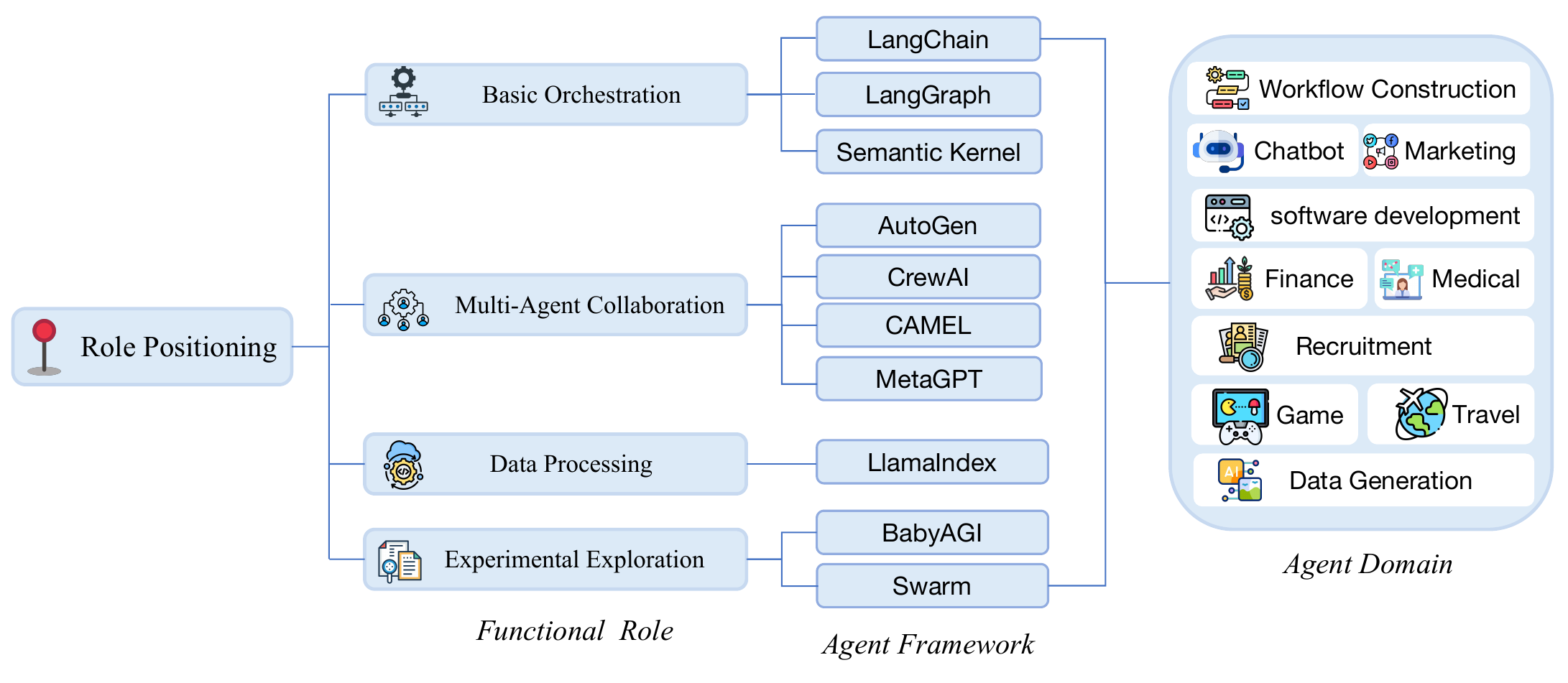}
    \vspace{-1em} 
    \caption{The Functional Roles and Domain of the Ten Agent Frameworks.}
    \label{fig:role}
    \vspace{-0.5em} 
\end{figure*}

\section{RQ1: LANDSCAPE OF FRAMEWORKS AND USAGE TRENDS}
\label{sec:RQ1}
To gain insights into the current state of development and evolutionary trends of agent frameworks, we conduct a preliminary analysis of 1,575 GitHub projects related to agent development, identifying ten widely adopted frameworks and examining their role positioning, usage patterns, and popularity trends from a developer's perspective.
\subsection{Analysis Method}
This section outlines the analytical methods used to understand the current state of LLM-based agent development frameworks. First, ten leading frameworks are identified through topic tag frequency analysis and manual relevance verification. Second, the functional positioning of ten frameworks is identified by classifying their core roles based on 8,710 GitHub developer discussion threads. Futhermore, co-occurrence analysis and dependency checks are conducted to examine their actual usage patterns in practice. 
\subsubsection{Framework Identification}

To identify the leading frameworks in LLM-based agent development, we follow the steps below:  \textbf{(i) Extracting topics.} We analyze the ``topics'' field of each project, which developers manually annotate to indicate the libraries, tools, and frameworks used in a given repository.  
\textbf{(ii) Ranking by frequency.} By calculating the frequency distribution of all topic tags in the dataset, we extract the ten most frequently occurring keywords, each corresponding to a distinct agent development framework.  
\textbf{(iii) Verifying relevance.} We ensure that these keywords indeed refer to agent frameworks rather than general-purpose dependencies or utility libraries, manually checking all relevant projects.  
\textbf{(iv) Removing false positives.} We refine the dataset by excluding projects that contain relevant tags but have no substantial connection to agent development.  
\textbf{(v) Collecting metadata.} Our dataset includes information on the number of stars, forks, used by, contributors and occurrences in 1,575 projects (repo counts) for the ten frameworks, collected up to July 2025.  
This process results in a representative list of widely used frameworks. It is worth noting that these ten frameworks do not encompass all existing frameworks; rather, they represent the top ten selected based on attention and relevance following our screening process.
\begin{figure*}[t]
    \centering
    \includegraphics[width=0.5\linewidth, keepaspectratio]{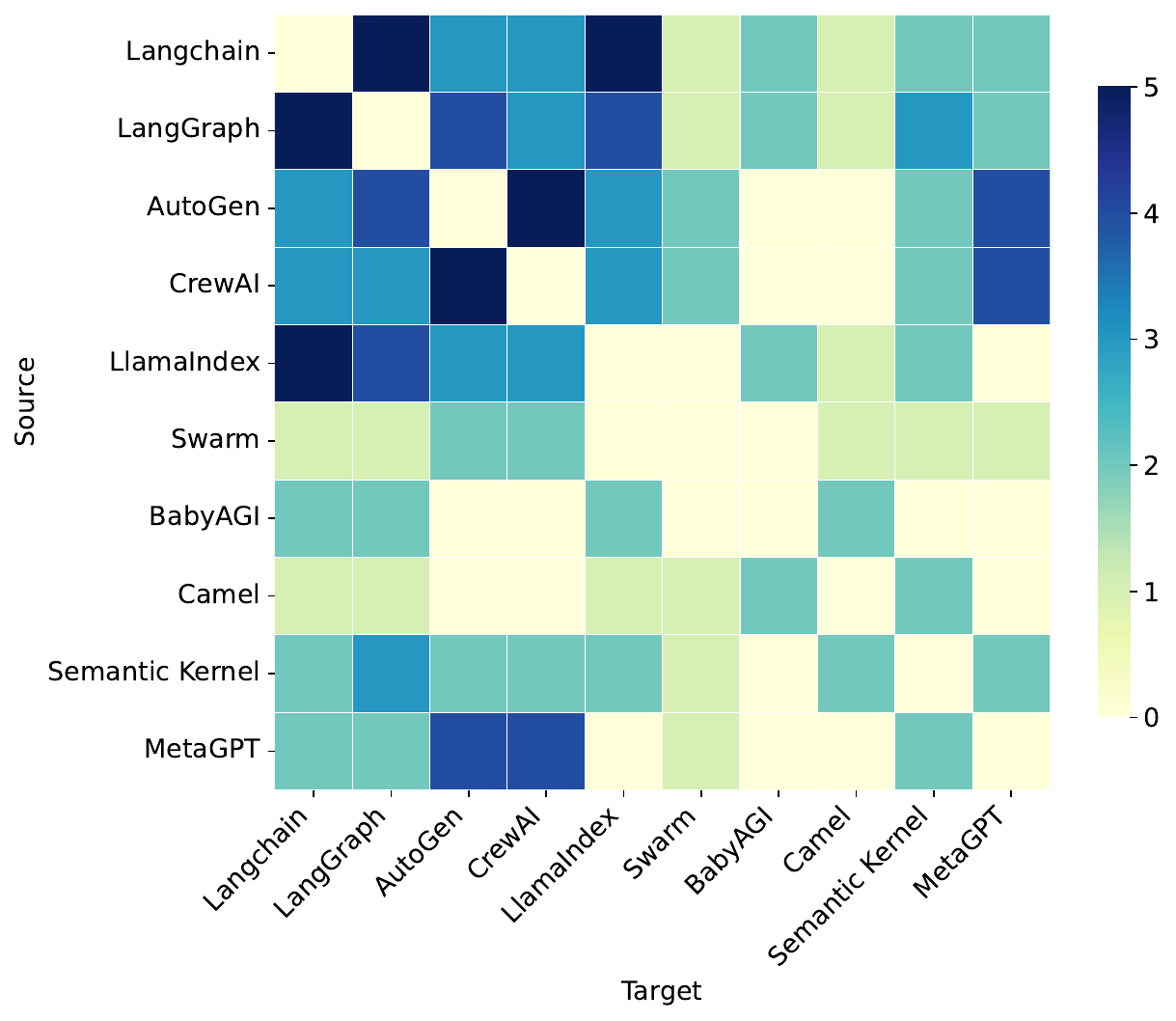}
    \vspace{-1em} 
    \caption{Heat Map of the Relationship Between the Combined Use of Ten Frameworks.}
    \label{fig:heat}
\end{figure*}
\subsubsection{Framework Roles}
To identify the functional roles of agent development frameworks in actual workflows, we analyze GitHub developer discussions. We do not rely solely on official documentation, but instead capture actual usage patterns and developer-centered insights derived from community discussions. Although most frameworks clearly specify their intended purposes in the documentation, these descriptions do not always align with actual usage. 

While GitHub discussion forums aggregate rich developer insights, their expression is inherently messy and heterogeneous. Developers may use different terms (\ie  ``tool invocation'' versus ``API orchestration'') to describe the same functionality or mention multiple frameworks within a single discussion thread, which poses challenges for direct classification. To address this problem, we develop a three-stage processing pipeline to extract functional roles from 8,710 developer discussion threads. 
\textbf{(i) Keyword extraction.} 
We first preprocess each discussion thread by tokenizing text and removing stop words. 
Then we apply the term frequency-inverse document frequency (TF-IDF) algorithm over the entire corpus to compute the importance score of each thread. 
For each thread, the top-ranked terms by TF--IDF weight are selected as representative keywords.
\textbf{(ii) Semantic classification.} We use the large language model GPT-4o to map each thread to a set of standardized functional tags, unifying different expressions into a cross-framework consistent semantic space.
\textbf{(iii) Frequency analysis.} We calculate the usage frequency of each functional tag within the discussion corpus of each framework to reveal usage patterns. For example, discussions about LangChain and AutoGen focus on task planning and tool integration, whereas LlamaIndex primarily relates to document indexing and retrieval-augmented generation.
\textbf{(iv) Expert review.} 
Two domain experts independently review all annotations to ensure classification accuracy and remove any discrepancies or ambiguities from the dataset. 
To quantify inter-rater reliability, we compute Cohen's~$\kappa$~ ~\cite{landis1977measurement} between the two annotators, obtaining a substantial agreement of $\kappa = 0.82$. 
All disagreements are removed from the final validated dataset.
Through this approach, we successfully transform unstructured developer discussions into structured functional categories and derive role-based classifications for 10 major agent development frameworks.

\subsubsection{Framework Usage}

We evaluate the frameworks' usage in actual development through both topic tags and dependency analysis.
\textbf{(i) Topics tag coverage.} We count the occurrence of framework-related topic keywords across all 1,575 projects to measure each framework's coverage within the repositories. Analyzing multi-tagged projects reveals emerging development practices adopted by developers when building agents. This approach assumes that co-occurring framework tags signal intentional integration rather than coincidence. By focusing on multi-tagged repositories, we surface patterns in how developers combine toolchains.
\textbf{(ii) Dependency verification.} To improve accuracy, we examine the repository structure of all 1,575 projects, focusing on \texttt{requirements.txt} and \texttt{package.json} files to check whether frameworks are explicitly declared as dependencies. Although this process is time-consuming, it allows us to verify whether the topics field accurately reflects actual framework usage.

\begin{table}[t]
\centering
\caption{Statistics of Ten Widely Used LLM-based Agent Frameworks. ``-'' means means that no relevant data is publicly available.}
\label{tab:framework_stats}
\begin{tabular}{lrrrrr}
\toprule
Agent Framework & \# Stars (k) & \#
Forks (k) & \#
Used~by & \#
Contributors & \# Repo~Counts \\
\midrule
LangChain        & 119  & 19.6 & 272k & 3798 & 105 \\
LangGraph        & 20.6  & 3.6  & 33.4k  & 269  & 26  \\
AutoGen          & 51.4 & 7.8    & 3.9k & 559  & 22  \\
CrewAI           & 40 & 5.3  & 17.2k & 278  & 19  \\
LlamaIndex       & 45.1 & 6.5  & 23k& 1713 & 12  \\
Swarm            & 20.6   & 2.2    &   -   & 15   & 5   \\
BabyAGI          & 21.9 & 2.8  & -   & 76   & 3   \\
Camel            & 14.7  & 1.6   & 163   & 182   & 3   \\
Semantic Kernel  & 26.6 & 4.3  & - & 421  & 2   \\
MetaGPT          & 59.2 & 7.2  & 119   & 113  & 2   \\
\bottomrule
\end{tabular}
\vspace{-1em}
\end{table}

\subsection{Results}

\subsubsection{Framework Roles}
To understand how AI agent frameworks support diverse application scenarios, we categorize them according to their functional roles.
From functional standpoint, agent frameworks can be classified into four categories, see Figure~\ref{fig:role} : basic orchestration, multi-agent collaboration, data processing, and experimental exploration. These categories collectively support a diverse range of application domains, including workflow automation, conversational agents, marketing, offline deployment, software engineering, finance, healthcare, recruitment, gaming, travel, and data generation. 

\textbf{Basic orchestration.} 
These frameworks provide interfaces, integrating models, and tools to support structured task management. Representative examples include LangChain, LangGraph, and Semantic Kernel. LangChain has achieved the widest adoption, particularly in chatbot construction and document analysis. LangGraph introduces a ``breakpoint'' mechanism that allows human review and approval at critical workflow nodes, which makes it particularly suited to high-stakes domains such as financial risk management and medical diagnostics. Semantic Kernel, in contrast, emphasizes the automation of traditional business workflows through a plugin-based architecture. 

\textbf{Multi-agent collaboration.} 
Multi-agent collaboration frameworks enable division of labor and problem solving among agents, typically through role definitions and interaction protocols. Notable frameworks include AutoGen, CrewAI, CAMEL, and MetaGPT. AutoGen is frequently applied in software development and other domains requiring multi-agent collaboration. CrewAI is oriented toward marketing, travel, recruitment, gaming, and finance. MetaGPT targets software development, while CAMEL leverages expert-role simulation to generate large-scale data and code.  

\textbf{Data processing frameworks.} They focus on knowledge augmentation, and information retrieval. LlamaIndex exemplifies this category and is widely used in chatbot construction and knowledge-base question answering.

\textbf{Experimental Exploration frameworks.} Experimental Exploration frameworks is served as testbeds that inform the development of subsequent frameworks. BabyAGI represents this category, offering a lightweight solution for automated task management. Swarm provides a lightweight platform for developers to experiment with multi-agent.

\begin{center}
\begin{tcolorbox}[colback=gray!10,
                  colframe=black,
                  width=\textwidth,
                  arc=1mm, auto outer arc,
                  boxrule=0.5pt,
                 ]
\noindent
\emph{\textbf{Finding 1:} The ten LLM-based agent frameworks serve functional roles in four categories, namely basic orchestration, multi-agent collaboration, data processing, and experimental exploration, and are applied across ten domains including software development.
}
\end{tcolorbox}
\end{center}

\subsubsection{Usage patterns}

To better identify the development patterns in AI agent engineering while using frameworks, we analyze the top $25\%$  starred Github projects to understand their actual framework usage. We find that $96\%$ of these projects employe two or more different agent-related frameworks, as shown in the figure \ref{fig:heat}. This high proportion indicates that a single-framework solution is no longer sufficient to meet the complex demands of real-world agent applications. 

We identify two widely used patterns. \textbf{The first involves orchestration frameworks combined with data frameworks}, with a typical example being LangChain paired with LlamaIndex. The former provides flexible workflow and task orchestration capabilities, while the latter supports efficient data indexing and retrieval. Their combination allows for complementary strengths, enabling more complex application scenarios.  \textbf{The second pattern involves multi-agent frameworks combined with orchestration frameworks}, commonly exemplified by AutoGen with LangChain. LangChain serves as a general orchestration layer, compatible with nearly all mainstream large language model APIs and hosting platforms. By using LangChain as AutoGen's model client, developers can seamlessly switch between models on platforms such as OpenAI, Anthropic, and Hugging Face without modifying AutoGen's core code, significantly enhancing system flexibility and scalability.

\begin{center}
\begin{tcolorbox}[colback=gray!10,
                  colframe=black,
                  width=\textwidth,
                  arc=1mm, auto outer arc,
                  boxrule=0.5pt,
                 ]
\noindent
\emph{\textbf{Finding 2:} $96\%$ of top-starred projects adopt multiple frameworks, highlighting that a single framework can no longer meet the complex needs of agent systems. Developers are combining multiple frameworks to build agents. 
}
\end{tcolorbox}
\end{center}

\begin{figure*}[t]
    \centering
    \includegraphics[width=0.7\linewidth, keepaspectratio]{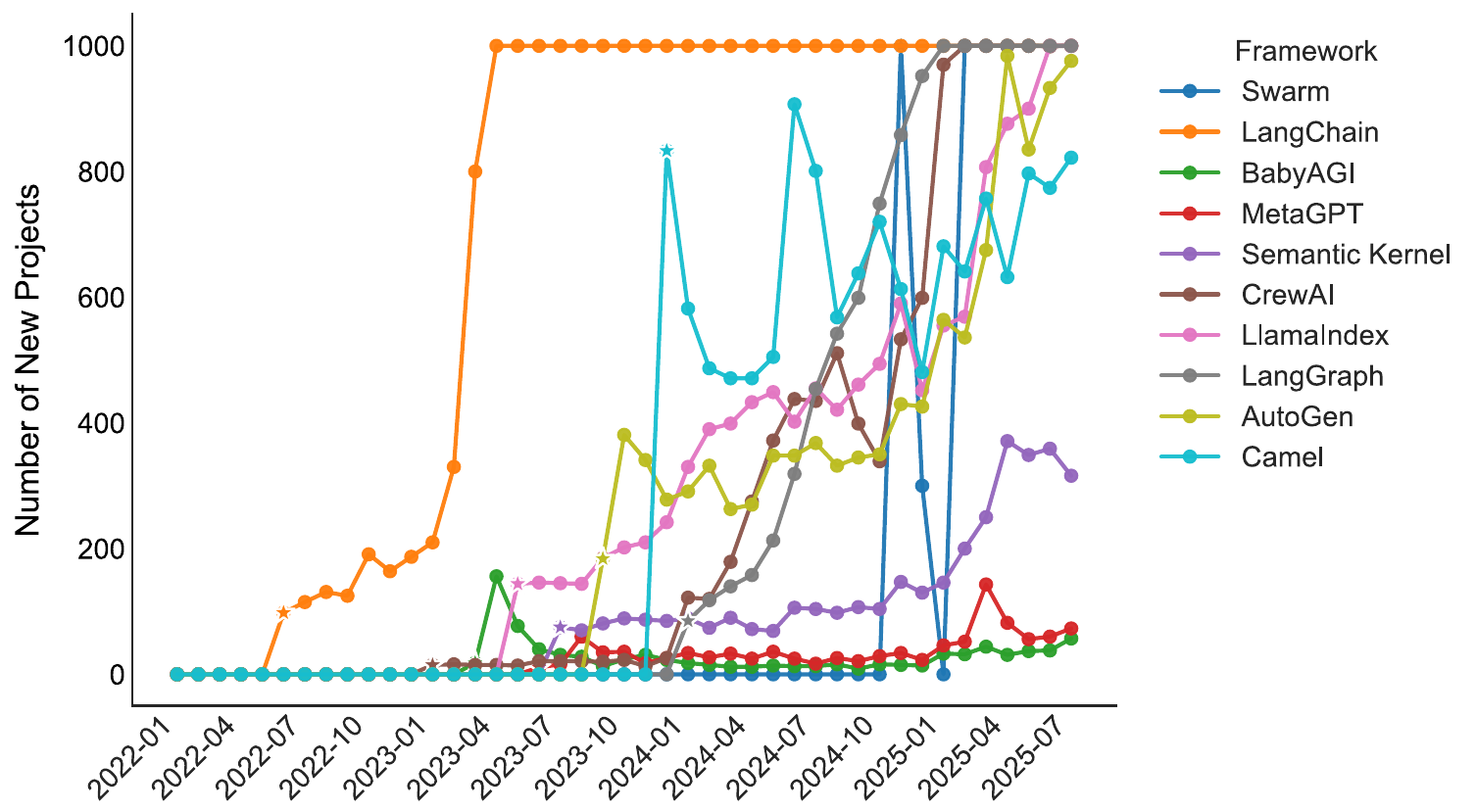}
    \vspace{-0.5em}
    \caption{Growth Trajectories of Github Projects Across the Ten Agent Frameworks.}
    \vspace{-1em}
    \label{fig:framework_trends}
\end{figure*}

\subsubsection{Community Popularity}

To gain community popularity and real-world adoption of the agent framework, we analyze multiple indicators, including GitHub Stars, Forks, the number of dependent projects (used by), contributor counts (Contributors), and actual adoption across 1,575 agent-related repositories (Repo Counts) see Table~\ref{tab:framework_stats}. Moreover, we examine the monthly usage trends of each framework from 2022 to 2025  (see Figure~\ref{fig:framework_trends}).

\textbf{High popularity but limited adoption.} Although certain frameworks have received significant attention on GitHub, they exhibit limited adoption in real-world development(See in Figure \ref{fig:framework_trends},\ref{tab:framework_stats}). For instance, MetaGPT has accumulated 48.7K \emph{Stars} and 5.8K \emph{Forks}, yet it is represented in only two repositories in our dataset. A plausible explanation is that MetaGPT is primarily designed as a demonstration project rather than a general-purpose framework, which limits its ecosystem consolidation and long-term sustainability. The same is true for other frameworks such as BabyAGI and Camel. Their monthly usage curves do not reflect sustained adoption, as these frameworks lack ongoing ecosystem support, limiting their long-term adoption.

\textbf{Low popularity but steadily growing adoption.}  
In contrast, LangGraph, despite having only 9.6K \emph{Stars}  and 1.2K Forks, is adopted in 26 repositories. Its monthly usage rose rapidly after early 2025, representing the second-highest adoption among the surveyed frameworks. 
Overall, the trends in Figure~\ref{fig:framework_trends} are consistent with the static data in Table~\ref{tab:framework_stats}, indicating that community popularity alone is not a reliable predictor of real-world framework adoption. Frameworks that achieve sustained use in development practice typically rely on ecosystem maturity, consistent maintenance, and targeted support for specific development needs. 

\begin{center}
\begin{tcolorbox}[colback=gray!10,
                  colframe=black,
                  width=\textwidth,
                  arc=1mm, auto outer arc,
                  boxrule=0.5pt,
                 ]
\noindent
\emph{\textbf{Finding 3:} The gap between a framework's community popularity and its actual real-world adoption suggests that developers should prioritize factors like ecosystem maturity and maintenance activity when selecting a framework, rather than relying solely on short-term indicators such as GitHub stars.      
}
\end{tcolorbox}
\end{center}

\UseRawInputEncoding
\pdfoutput=1
\section{RQ2: CHALLENGES ACROSS THE SOFTWARE DEVELOPMENT LIFECYCLE }
\begin{figure*}[t]
    \centering
    \includegraphics[width=0.9\linewidth, keepaspectratio]{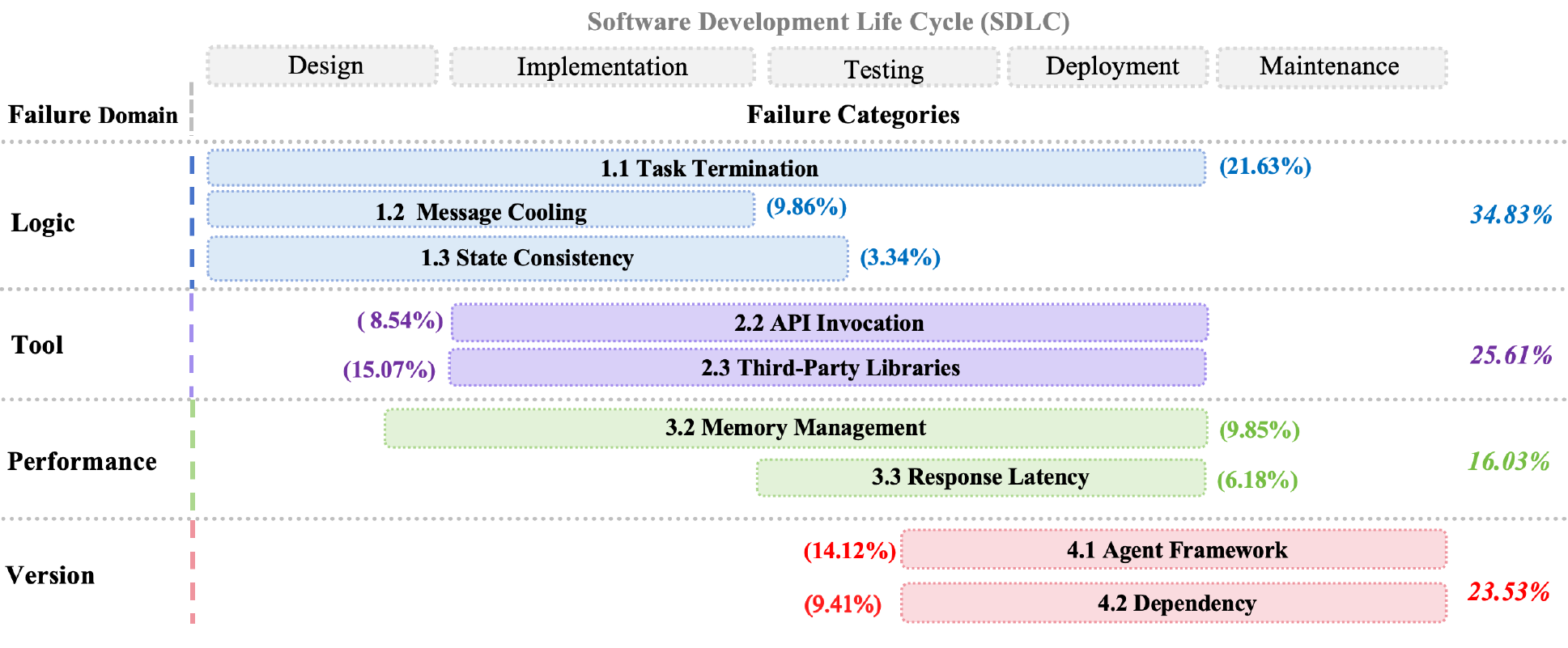}
    \caption{AI Agent Framework Failure Analysis Across the Software Development Life Cycle.}
    \label{fig:RQ2}
\end{figure*}

Based on RQ1, we have identified the functional roles and usage patterns of ten major agent frameworks. To further investigate the challenges of these frameworks that hinder productivity and compromise system stability in real-world agent development, in this section, we present the results of our analysis of developer challenges in LLM-based agent development. We map 8,710 LLM-generated issue summaries to the five phases of the Software Development Lifecycle (SDLC) to capture when problems occur. Based on cross-phase clustering and expert review, we further identify four overarching challenge categories: \textit{Logic}, \textit{Tool}, \textit{Performance}, and \textit{Version}, which describe what types of problems developers face.
Together, this two-dimensional framework (SDLC $\times$ challenge category, see in figure~\ref{fig:RQ2}) enables both temporal and structural analysis of agent development difficulties. The proportions of each category indicate their relative prevalence and severity across the lifecycle, providing a quantitative reference for prioritizing framework improvements and developer support mechanisms.
\label{sec:RQ2}
\subsection{Analysis Method}
\label{sec:3.2.1}

This section outlines the methods used to summarize and categorize developer discussions.
To begin with, multi-turn GitHub threads are transformed into structured inputs enriched with contextual metadata (project, framework, date, author role).
Next, developer intent is extracted through carefully designed system and user prompts.
Finally, GPT-4 performs automated inductive summarization, generating concise one-sentence JSON outputs.
Together, these summaries provide the basis for mapping issues to SDLC stages and for systematically analyzing the key challenges in LLM-based agent development.
\subsubsection{Discussion Summarization Via LLM}
\label{sec:3.2.1.1}
In this subsection, our goal is to distill the multi-turn, lengthy, and heterogeneous expressions found in raw GitHub discussions into concise representations of developer intent. These summaries serve as the foundation for subsequent mapping to software lifecycle and issue categories. To generate concise developer issue summaries, we adopt a three-step approach:  
\textbf{(i) Constructing inputs.} We build inputs from the 8,710 cleaned discussion threads. Each thread is formatted to include the discussion title and main content, along with contextual metadata such as project name, framework, creation date, and author role.  
\textbf{(ii) Intent extraction.} To achieve high-quality extraction of developer intent, we adopt dedicated system prompts and user prompts ~\cite{white2023prompt}.  
\textbf{(iii) Automated summarization.} We use GPT-4 for inductive summarization, and the entire process is fully automated.

\begin{figure}[ht]
    \centering
    \vspace{-10pt}

    \begin{tcolorbox}[breakable,
        title=\textbf{Prompts Used in Discussion Summarization},
        colback=white!95,
        colframe=gray!70!black,
        fonttitle=\bfseries,
    ]

        \textbf{System Prompt.} 
        I want you to act as a data scientist to analyze datasets. 

        Do not make up information that is not in the dataset. 

        For each analysis I ask for, provide me with the exact and definitive answer 
        and do not provide me with code or instructions to do the analysis on other platforms.

        \tcblower %

        \textbf{User Prompt.}
        Please strictly follow the instructions below:

        1. Identify information related to the agent development process.\\
        2. Summarize the inferred information into a single concise sentence.\\
        3. Count the frequency of each information category.\\
        4. Output in pure JSON format only, with no explanations.

        Data to process: \{\textbf{batch\_text}\}

        Ensure that the output strictly conforms to the required JSON format.

    \end{tcolorbox}
    \vspace{-5pt}
    \caption{System and User Prompts Used in Discussion Summarization.}
    \label{fig:discussion-prompts}
\end{figure}

With this prompt pair ~\cite{schulhoff2024prompt}, the model extracte the core issue discussed in each thread and generated a concise one-sentence summary in a standardized format. Each result is returned in a clear JSON structure, thereby simplifying subsequent analysis and topic classification.

\subsubsection{Issue Categorization In Software Development Lifecycle}
To better understand when and where specific challenges arise during AI agent development, we construct a taxonomy of issues based on the Software Development Lifecycle (SDLC). SDLC is widely used in traditional software engineering to describe the phased progression of system development, from design to deployment and maintenance, we adapt SDLC to reflect the unique workflows and pain points observed in LLM-based agent development. Our taxonomy involves two key steps:  
\textbf{(i) Define stage-issue mapping.} We first map common agent development tasks to five SDLC stages: design, implementation, testing, deployment, and maintenance. For each stage, we outline the main developer activities and infer the types of technical issues that occur. In this paper, we merge requirements analysis into the design stage, as the two are often closely intertwined in agent development, with requirements directly shaping system architecture and design decisions.  
\textbf{(ii) Annotate discussions by stage.} We use the LLM-generated issue summaries (as described in Section ~\ref{sec:3.2.1.1}) as the basis for annotating SDLC stages. Each summary is reviewed to identify contextual cues indicating where the issue falls within the development timeline. For example, ``difficulties in configuring multi-agent workflows on Docker'' is classified under the deployment stage. To ensure classification quality, we manually review 500 issue summaries.

\begin{center}
\begin{tcolorbox}[colback=gray!10,
                  colframe=black,
                  width=\textwidth,
                  arc=1mm, auto outer arc,
                  boxrule=0.5pt,
                 ]
\noindent
\emph{\textbf{Finding 4:} The challenges faced by LLM developers are multifaceted, covering 4 domain and 9 distinct categories.
}
\end{tcolorbox}
\end{center}
\subsection{Results}

\begin{figure*}[t]
    \centering
    \includegraphics[width=0.9\linewidth, keepaspectratio]{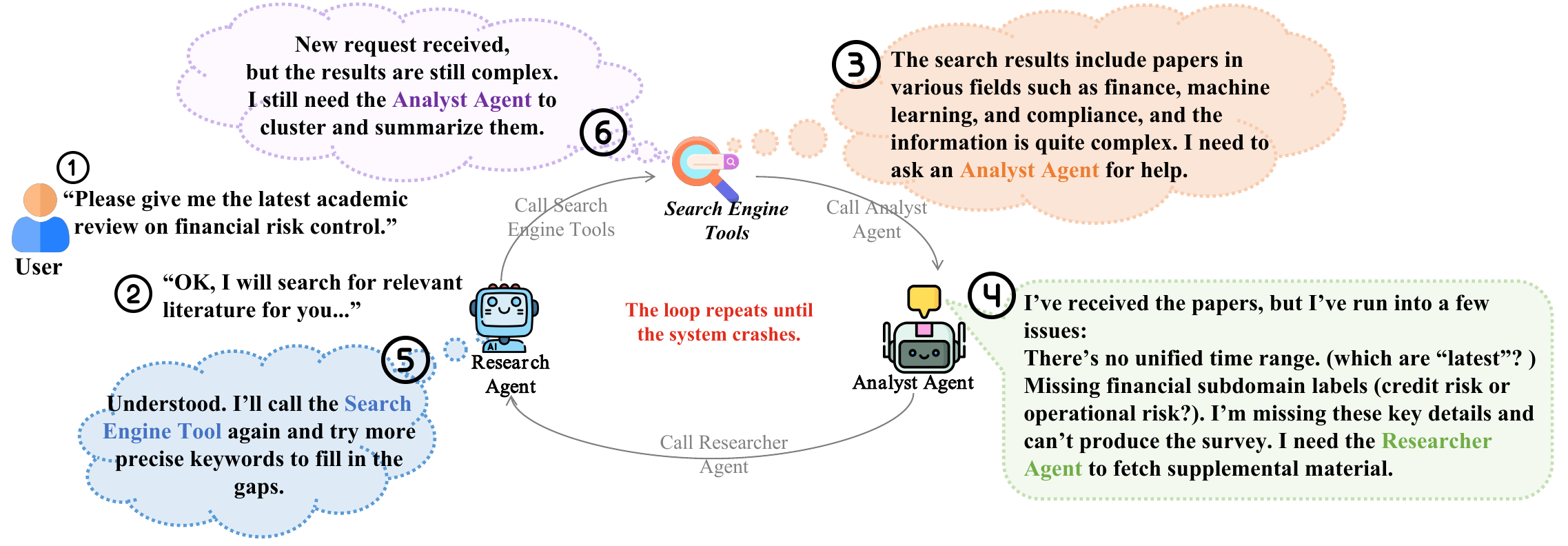}
    \caption{The Agent Falls Into an Infinite Loop When Executing Tasks.}
    \label{fig:finding4}
    \vspace{-0.5em} 
\end{figure*}

\textbf{Logic Failures.}
Logic-related failures summarize those faults that originate from deficiencies in the agent's internal logic control mechanisms. Typical cases include task termination failures, context misalignment, and state inconsistency. Their common characteristic lies in flaws of decision-making and scheduling logic design. Task termination issues account for more than 21\% of all observed failures. \blackcircled{1} \textbf{The lack of robust termination mechanisms} leads to task flows failing to complete properly (See in Figure\ref{fig:finding4}).  About 8\% of cases exhibit recursive tool or agent calls, which continue execution until the system enforces an iteration limit.  72\% of these recursive failures occur at the interaction layer between the agent and external tools. The root causes include missing call-chain state tracking, insufficient dynamic termination detection, and the lack of rollback strategies. In high-concurrency scenarios, these loops are even harder to interrupt externally, severely affecting system stability and user experience. Therefore, designers of agent frameworks need to define explicit task coordination mechanisms for multi-agent frameworks.
\blackcircled{2} \textbf{Insufficient message management} causes a large number of inefficient redundant behaviors. 64\% of redundant interactions stem from the lack of message cooling mechanisms and relevance checks, with agents frequently triggering duplicate responses in shared dialogue spaces, or continuing to pass along irrelevant context after a task is finished, resulting in ``ineffective dialogue loops.'' This behavior not only wastes computational resources and increases API costs, but also causes delays and rate-limit triggers during long-running interactions.

\begin{center}
\begin{tcolorbox}[colback=gray!10,
                  colframe=black,
                  width=\textwidth,
                  arc=1mm, auto outer arc,
                  boxrule=0.5pt,
                 ]
\noindent
\emph{\textbf{Finding 5:} More than one-third of the failures are caused by internal logic control deficiencies. 
Among them, task termination and message cooling issues account for 21.63\% and 9.86\%, respectively, 
highlighting the importance of termination control and message management.
}
\end{tcolorbox}
\end{center}
\textbf{Tool Failure.}
Tool integration issues occur widely across the design, development, testing, and deployment stages. Agent-based systems particularly those involving large models and third-party dependencies often face deployment resistance. \textbf{API integration and third-party service interfacing are core pain points in the implementation phase}, with approximately $14\%$ of reported issues related to integration and adaptation barriers, such as API rate limits, permission errors, and missing dynamic libraries. \textbf{The root cause lies in the lack of standardized API design and guarantees of cross-framework compatibility}, which results in errors or degraded performance when integrating external tools or services into agent frameworks. Representative cases include the discussion on the ``AzureOpenAI client failing to define a complete URL''. Similar integration failures with tools such as Langfuse and OpenTelemetry are also common, typically caused by missing modules, or configuration errors. These problems slow down the process of agent development. The introduction of the Model Context Protocol (MCP) has partially improved cross-tool and cross-service compatibility, but it has not fully resolved the issue.

\begin{figure*}[t]
    \centering
    \includegraphics[width=0.6\linewidth, keepaspectratio]{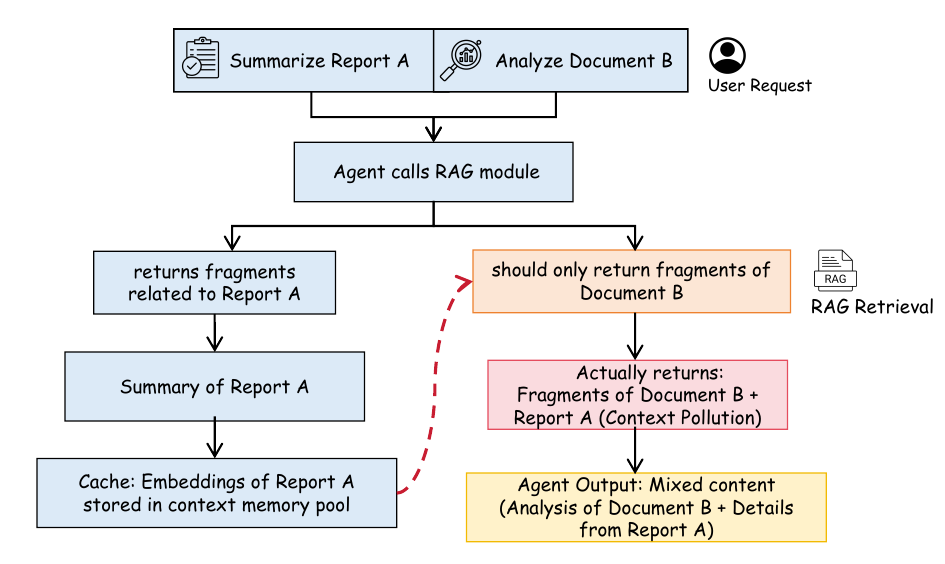}
    \caption{Agent Memory Management Failure Leads to Confusion of Session History.}
    \label{fig:finding6}
    \vspace{-0.1em} 
\end{figure*}

\begin{center}
\begin{tcolorbox}[colback=gray!10,
                  colframe=black,
                  width=\textwidth,
                  arc=1mm, auto outer arc,
                  boxrule=0.5pt,
                 ]
\noindent
\emph{\textbf{Finding 6:} API integration and connecting to third-party services are major challenges. $25.61\%$ of issues involve API limitations, permission errors, and missing dynamic libraries. 
}
\end{tcolorbox}
\end{center}
\textbf{Performance Failure.}
Performance safety issues account for 16.03\% of all identified failures, primarily arising from deficiencies in memory management ( See in \ref{fig:finding6}) and response latency.
\blackcircled{1} \textbf{Memory management.} When conversations exceed 20 turns, some frameworks experience fragmented responses. This issue is primarily due to inadequate memory caching strategies. When multiple threads simultaneously call the same knowledge index, state pointers can be overwritten, causing memory vectors to be lost between conversations. Furthermore, some frameworks' temporary caches default to local memory rather than persistent storage, resulting in complete loss of historical context after system restarts.
\blackcircled{2} \textbf{Response latency.} Latency remains a bottleneck in many agent frameworks, especially those relying on synchronous message passing or  retrieval-augmentation pipelines. For example, LlamaIndex and LangGraph typically perform multi-stage retrieval and embedding operations before generating a response, introducing waiting time between retrieval and reasoning phases. The average end-to-end latency of retrieval-augmented agents ranges between 3.2 s and 5.6 s per query, approximately 1.8 slower than direct generation pipelines on comparable hardware (A100 GPU, batch = 1)

\begin{figure*}[t]
    \centering
    \includegraphics[width=0.9\linewidth, keepaspectratio]{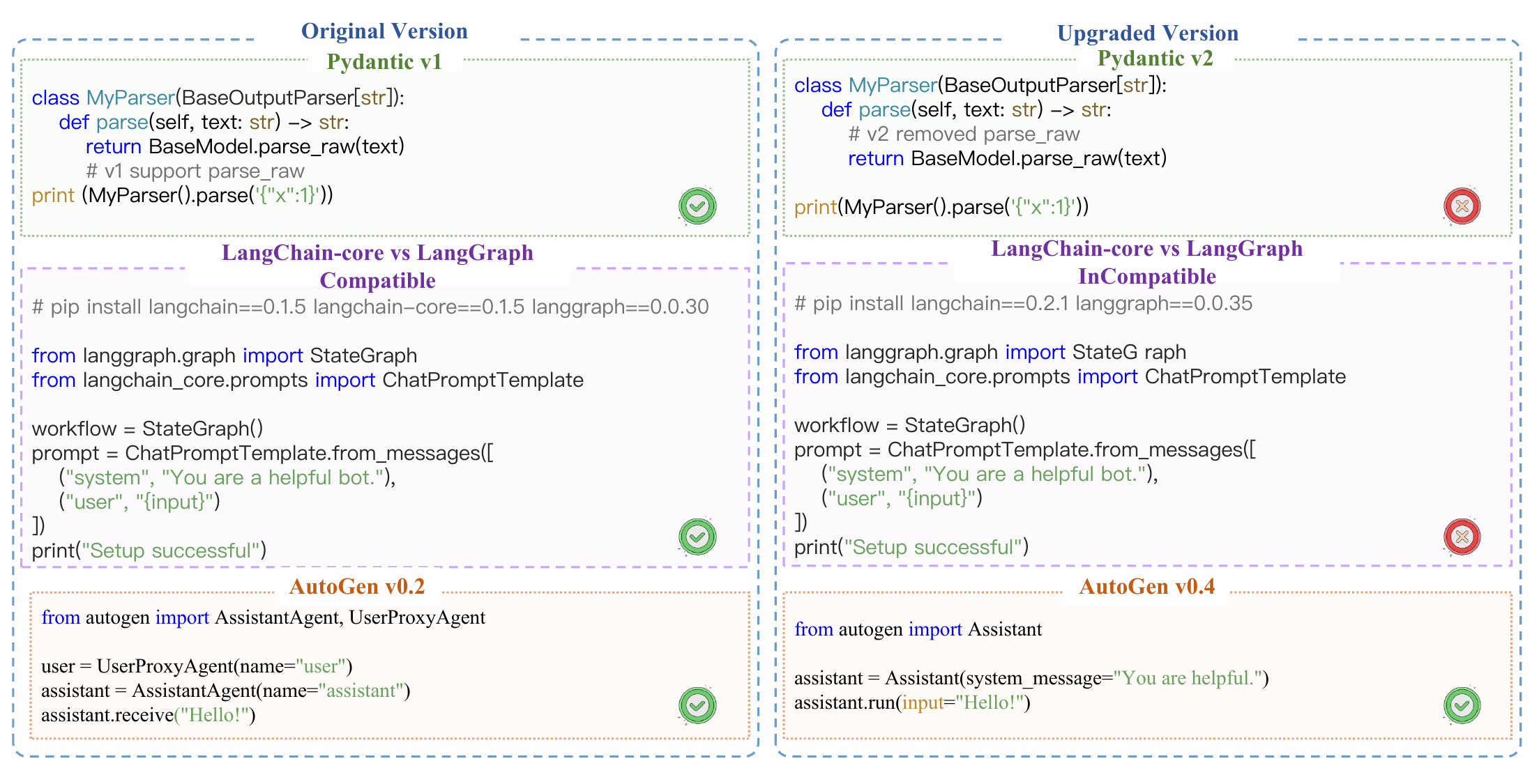}
    \caption{Illustrative Examples of Dependency-Induced Failures in Agent Frameworks.}
    \label{fig:finding7}
    \vspace{-0.1em} %
\end{figure*}

\begin{center}
\begin{tcolorbox}[colback=gray!10,
                  colframe=black,
                  width=\textwidth,
                  arc=1mm, auto outer arc,
                  boxrule=0.5pt,
                 ]
\noindent
\emph{\textbf{Finding 7:} Performance issues, accounting for 16.03\% of all identified failures, predominantly stem fromunstable memory persistence, and high response latency.
}
\end{tcolorbox}
\end{center}
\textbf{Compatibility Failure.}
In the AI agent ecosystem, version conflicts have become the direct cause of over $25\%$ of technical obstacles, occurring most frequently during the deployment and maintenance stages. Agent systems are composed of rapidly evolving components such as large language models, vector databases, external API tools, and orchestration frameworks. This ``glue-like'' architecture is highly vulnerable to chain reactions triggered by updates to individual components, resulting in functionality breakdowns or even system crashes. Typical examples( See in \ref{fig:finding7})include large-scale build failures in  \blackcircled{1} \emph{LangChain during the migration from Pydantic v1 to v2} due to breaking changes, and installation conflicts caused by inconsistent version requirements between internal components like  \blackcircled{2} \emph{langchain-core and langgraph}. An oversized dependency tree means that any update to a sub-dependency can destabilize the existing environment. CrewAI's strict version pinning of underlying libraries such as LiteLLM and Pydantic frequently leads to incompatibility in multi-framework collaboration.  \blackcircled{3} \emph{AutoGen's refactoring from v0.2 to v0.4} introduced a completely incompatible new architecture, which also split the community and resulted in multiple PyPI packages, creating uncertainty in adoption and subsequent maintenance. These problems not only increase development and maintenance costs but also force teams to invest significant effort in environment management to mitigate the systemic risks posed by version dependencies.
\begin{center}
\begin{tcolorbox}[colback=gray!10,
                  colframe=black,
                  width=\textwidth,
                  arc=1mm, auto outer arc,
                  boxrule=0.5pt,
                 ]
\noindent
\emph{\textbf{Finding 8:} The agent framework suffers from version compatibility traps, and $23.53\%$ of technical obstacles are directly related to version dependencies. 
}
\end{tcolorbox}
\end{center}

\UseRawInputEncoding
\pdfoutput=1
\section{RQ3: COMPARISON OF FRAMEWORKS IN MEETING DEVELOPER NEEDS}
\label{sec:RQ3}

In RQ2, we analyze the challenges that agent developers encounter when using agent frameworks during the development process. However, it remains unclear how existing frameworks address these challenges and whether they truly meet developers' needs. Therefore, based on the key challenges identified in RQ2 and the needs reflected in developer discussions, we propose five evaluation dimensions to assess how the ten frameworks identified in RQ1 support developers across different stages of agent development.

\subsection{Analysis Method}

\begin{table}[t]
\centering
\caption{Developer Evaluation Dimensions of Frameworks, With Definitions, Representative Examples From Discussions, and Their Proportions.}
\resizebox{0.95\textwidth}{!}{%
\begin{tabular}{@{}>{\raggedright\arraybackslash}m{3cm} 
                >{\raggedright\arraybackslash}m{4.5cm} 
                >{\raggedright\arraybackslash}m{6.5cm} 
                >{\centering\arraybackslash}m{2cm}@{}}
\toprule
\textbf{Dimension} & \textbf{Definition} & \textbf{Example from Discussions} & \textbf{Proportion} \\ \midrule
\makecell[l]{Learning \\ Cost} & The learning threshold of a framework, including the ease of documentation, tutorials, and community support & ``The documentation is too brief, I still don't understand after reading for a long time...'' & 18.70\% \\
\midrule
\makecell[l]{Functional \\ Abstraction} & Whether the framework provides clear abstractions and componentized design to simplify the invocation of complex functions & ``This is such a common requirement, but I have to reinvent the wheel... lacking abstraction.'' & 21.10\% \\
\midrule
\makecell[l]{Development \\ Efficiency} & The extent to which the framework helps developers reduce repetitive work and improve development speed & ``Every debugging session takes a huge amount of time, overall efficiency feels too low.'' & 28.40\% \\
\midrule
Maintainability & Stability, extensibility, and code readability of the framework in long-term use & ``After a few months, I don't dare to touch it again... too difficult to maintain.'' & 14.50\% \\
\midrule
\makecell[l]{Performance \\ Optimization} & Performance of the framework in terms of execution efficiency, memory usage, and task processing speed & ``I found the response very slow, with almost no optimization methods...'' & 12.30\% \\ 
\bottomrule
\end{tabular}%
}
\vspace{-1em}
\label{tab:framework_dimensions}
\end{table}

Our analytical method follows a multi-step process.  
\textbf{(i) Reviewing discussions.} We carefully read each collected discussion thread in full, with particular attention to the context in which developers raise concerns, highlight strengths, or describe trade-offs.  
\textbf{(ii) Open Coding and categorization.} 
We employ an open coding procedure, in which textual excerpts are labeled with descriptive codes that capture their core meaning. 
To ensure the reliability of the coding process, two researchers independently code a randomly selected subset of the data and compute inter-rater reliability using Cohen's~$\kappa$~ ~\cite{landis1977measurement}, obtaining $\kappa = 0.81$.
\textbf{(iii) Mapping to evaluation dimensions.}  
The themes categorized in the previous stage are systematically mapped to the five evaluation dimensions defined in Table~\ref{tab:framework_dimensions}.
\textbf{(iv) Quantitative analysis.} Instead of relying solely on qualitative interpretation, we count the frequency of mapped instances within each dimension. This produces a distribution table across the five dimensions, which serves as the empirical basis for our comparative evaluation.  

This stepwise mapping procedure provides a transparent and reproducible foundation for analyzing how developers perceive the strengths and weaknesses of different frameworks throughout the software development lifecycle.

\subsection{Result}
\subsubsection{Learning Cost}
Different frameworks vary significantly in terms of technical barriers to entry, documentation, community support, and the richness of examples. AutoGen reduces the entry barrier, offering clear and responsive documentation and over 200 reproducible notebook examples covering fields such as finance and healthcare. However, the cost can rise sharply later on, as it requires fine-grained control of conversation dynamics and an understanding of the underlying asynchronous event-driven architecture.
Despite its high level of abstraction and steep learning curve, LangChain, with its documentation, large community, and over 500 full-scenario examples, \textbf{\textit{is suitable for developers seeking flexibility and customization.}}  LlamaIndex and LangGraph are technically intuitive, but their documentation is fragmented or relies on other frameworks. Semantic Kernel's language features are inconsistent, plugin management is complex, and many examples are outdated. Frameworks with high barriers to entry and high learning costs include MetaGPT, which has frequent architectural changes and limited cross-platform compatibility. Camel has abstract examples and lacks practical tool integration; and BabyAGI, which requires manual handling of agent loops and state persistence while relying on outdated documentation.
CrewAI's declarative task orchestration is suitable for beginners, but its documentation is insufficient. The Swarm module, which has the steepest learning curve, has confusing dependencies, insufficient community support, and difficult-to-run examples.

\begin{center}
\begin{tcolorbox}[colback=gray!10,
                  colframe=black,
                  width=\textwidth,
                  arc=1mm, auto outer arc,
                  boxrule=0.5pt,
                 ]
\noindent
\emph{\textbf{Finding 9:} LangChain and CrewAI lower the technical threshold for beginners, while LangChain and CrewAI provide excellent documentation, strong community support, and abundant examples to help developers solve problems quickly. 
}
\end{tcolorbox}
\end{center}

\begin{figure*}[t]
    \vspace{-1em} %
    \centering
    \includegraphics[width=0.9\linewidth, keepaspectratio]{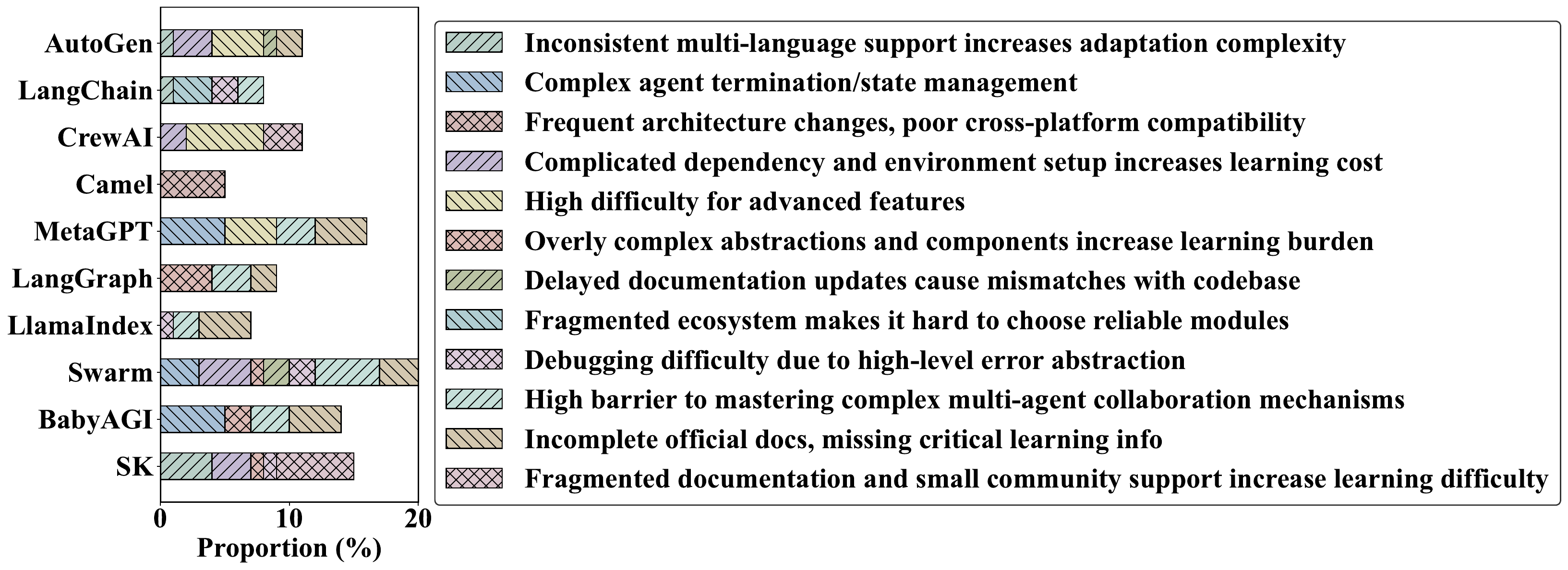}
    \caption{Various Issues Arising From the Cost of Learning. SK Denote Semantic Kernel.}
    \label{fig:overview}
    \vspace{-1em} %
\end{figure*}
\subsubsection{Development Efficiency}

AutoGen and LangChain demonstrate advantages in the rapid prototyping phase.  \textbf{Leveraging high-level abstractions and modular design, they lower the barrier to entry} for building LLM applications from scratch. In discussions, over $78\%$ of developers cited these two frameworks as enabling rapid prototype verification. Specifically, LangChain relies on a highly modular architecture, breaking down complex LLM workflows into pluggable components. This allows core functionality, such as search enhancement generation, to be implemented with just a few dozen lines of code. However, \textbf{while this layered encapsulation supports a wide range of scenarios, it also introduces additional cognitive burden}. Approximately $42\%$ of developers noted that \textbf{LangChain's deeply nested abstractions hindered development efficiency when dealing with complex or non-standard requirements}, with some even describing situations where ``a single change required traversing seven layers of code structure.'' AutoGen abstracts multi-agent collaboration into ``conversation,'' significantly lowering the barrier to entry for multi-agent development through role definition and automated interaction. However, \textbf{AutoGen suffers from tool compatibility deficiencies}. Approximately $31\%$ of projects experience adaptation failures or call errors when integrating custom tools. While these issues aren't fatal during rapid development, they can reduce development efficiency in complex scenarios and production environments. Semantic Kernel and LlamaIndex incur substantial debugging overhead, as developers need to spend considerable time locating and fixing errors, which in turn reduces overall development efficiency. BabyAGI is suitable for lightweight or task-specific experiments, but lacks support for distribution and containerization, limiting its scalability.

\begin{center}
\begin{tcolorbox}[colback=gray!10,
                  colframe=black,
                  width=\textwidth,
                  arc=1mm, auto outer arc,
                  boxrule=0.5pt,
                 ]
\noindent
\emph{\textbf{Finding 10:} AutoGen and LangChain excel at rapid prototyping, and their modular design reduces redundant code; however, LangChain's excessive abstraction and AutoGen's tool compatibility issues increase development difficulty. 
}
\end{tcolorbox}
\end{center}

\begin{figure*}[t]
    \centering
    \includegraphics[width=0.9\linewidth, keepaspectratio]{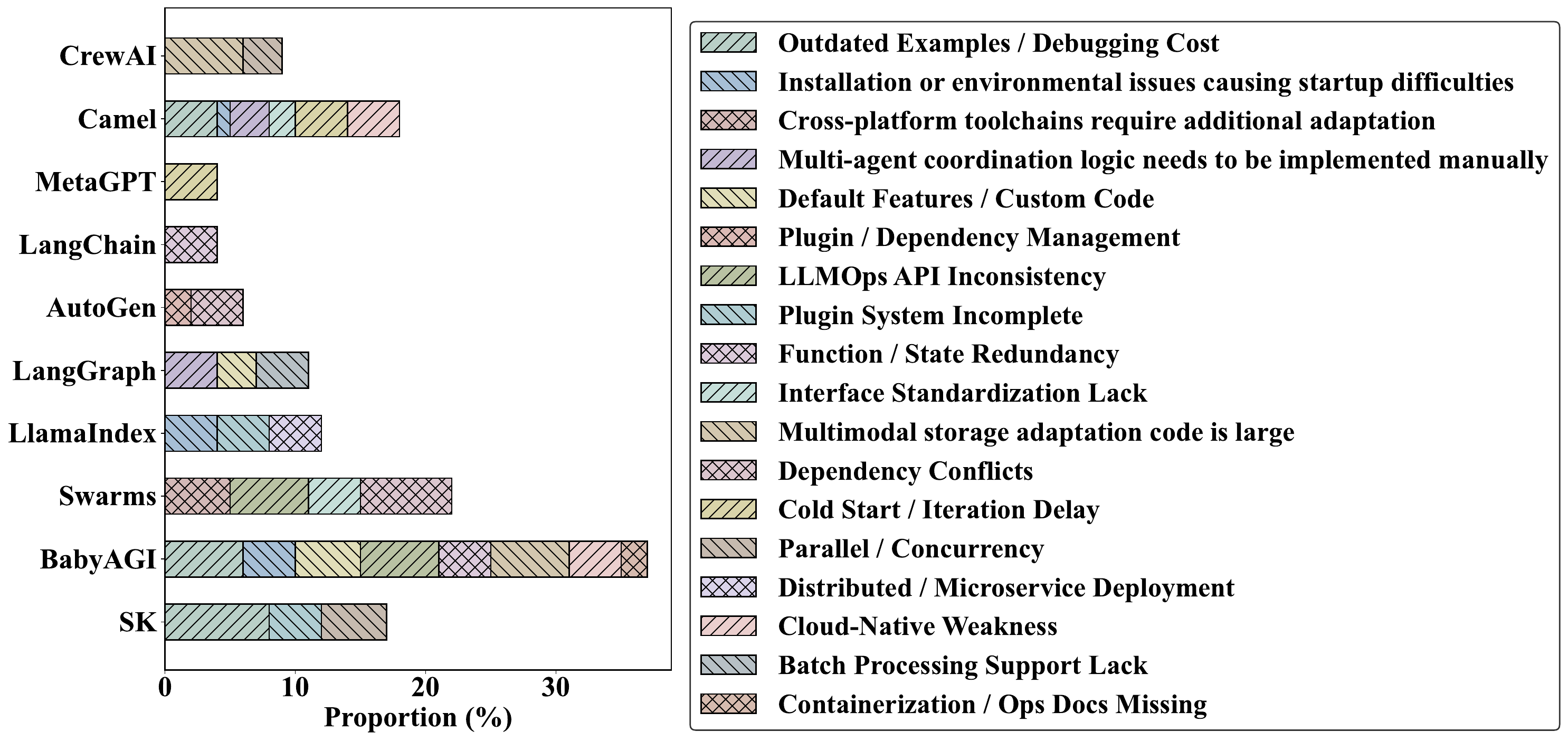}
    \caption{Various Issues Arising From Development Efficiency. SK Denote Semantic Kernel.}
    \label{fig:overview}
\end{figure*}
\subsubsection{Functional Abstraction}
At the functional abstraction level, frameworks differ significantly in task decomposition, context management, and multi-agent collaboration. AutoGen and LangChain excel in task decomposition and collaboration. \textbf{The former leverages the Studio visual designer and conversation-driven model for rapid prototyping}, but lacks multi-tenant isolation and is constrained by group chat routing under high concurrency. \textbf{The latter utilizes LCEL for chained composition, which offers high efficiency in some scenarios}s, but its excessive abstraction increases debugging costs for complex tasks. MetaGPT promotes task decomposition through code generation, facilitating business logic automation. However, the decomposition chain is susceptible to failure due to RAG deviation. LangGraph excels at visual orchestration, but lacks support for dynamic workflows and load balancing, reducing flexibility. CrewAI's hierarchical task templates are suitable for process-driven scenarios, but lack parameter validation and manual callbacks, reducing robustness. BabyAGI supports hierarchical delegation for lightweight use cases, but requires manual termination logic and lacks distributed synchronization, impacting the stability of long processes. Camel excels at simple interactions, but complex tasks rely heavily on manual coding and are not optimized for long contexts. Swarm lacks modules, complex multimodal setups, and is prone to interruption and lacks reliable failover. Semantic Kernel and LlamaIndex provide basic context persistence and plugin management, but task decomposition still relies on manual labor, resulting in insufficient context consistency. In particular, LlamaIndex lacks history management and conflict resolution, limiting the efficiency of multi-agent collaboration.

\begin{center}
\begin{tcolorbox}[colback=gray!10,
                  colframe=black,
                  width=\textwidth,
                  arc=1mm, auto outer arc,
                  boxrule=0.5pt,
                 ]
\noindent
\emph{\textbf{Finding 11:} In terms of functional encapsulation, AutoGen and LangChain are leading in task decomposition and multi-agent collaboration, but context consistency and high concurrency performance bottlenecks limit their production scalability.  
}
\end{tcolorbox}
\end{center}

\begin{figure*}[t]
    \centering
    \includegraphics[width=0.8\linewidth, keepaspectratio]{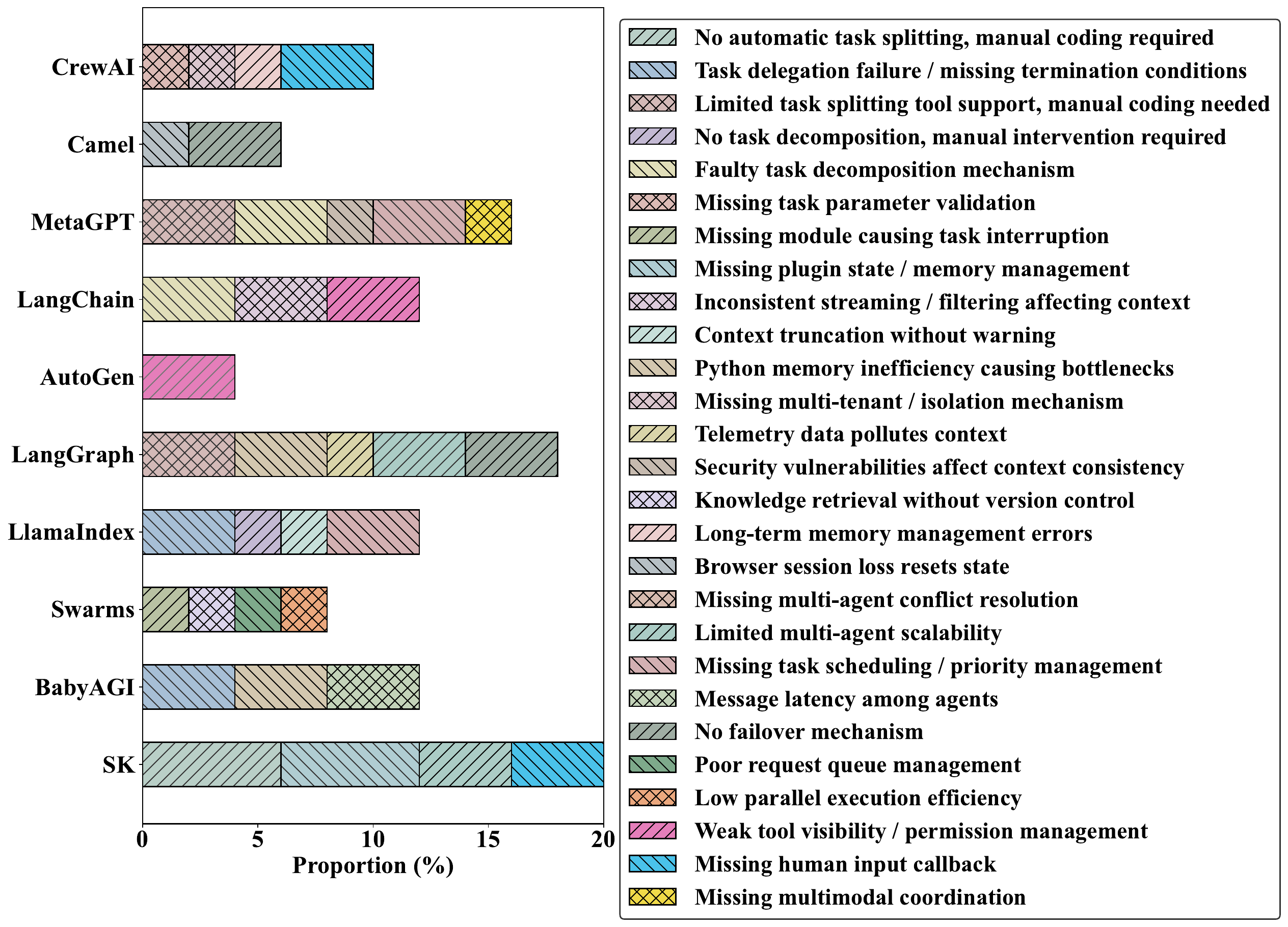}
    \caption{Various Issues Arising in Functional Abstraction. SK Denote Semantic Kernel.}
    \label{fig:overview}
    \vspace{-0.5em} %
\end{figure*}

\subsubsection{Performance Optimization }

Performance optimization is a common shortcoming across all frameworks. Frameworks differ in caching mechanisms, concurrent processing, and resource management. Semantic Kernellacks a built-in caching system, suffers from memory leaks, affects resource reuse, and presents performance bottlenecks in parallel execution and streaming token tracking. LlamaIndex has limited caching capabilities, ``MemoryStore'' migration and hybrid search operations lack efficient caching, parallel function calls frequently fail, and complex data types and vector storage updates lead to excessive memory usage. 

LangChain requires manual caching of batch operations for vector storage, lacks intermediate result caching in streaming processing, suffers from poor parallel aggregation performance, and generates high memory overhead for plugin management. LangGraph also lacks a caching mechanism. MetaGPT suffers from cache invalidation due to RAG index errors and does not provide query result caching. CrewAI suffers from inconsistent memory paths, no built-in caching API, rate limits affecting concurrent flows, and dependency conflicts leading to increased resource consumption. Swarms lacks a knowledge retrieval cache layer, requiring repeated reloading of multimodal data and frequently failing during parallel execution. AutoGen does not cache function call results or intermediate flow states, and suffers from low throughput for multi-agent messaging and tool calls.

\begin{center}
\begin{tcolorbox}[colback=gray!10,
                  colframe=black,
                  width=\textwidth,
                  arc=1mm, auto outer arc,
                  boxrule=0.5pt,
                 ]
\noindent
\emph{\textbf{Finding 12:} AutoGen, LangChain, and LangGraph face challenges with scalability and concurrent processing under high load or multi-agent scenarios, while SemanticKernel, LlamaIndex, BabyAGI, MetaGPT, CrewAI, Swarms, and Camel are limited by inefficient caching and insufficient resource management. These factors collectively hinder the performance of large-scale and complex workflows. 
}
\end{tcolorbox}
\end{center}

\subsubsection{Maintainability}
Frameworks differ in version upgrade support, long-term maintenance costs, and team collaboration standards. \textbf{\textit{Despite their mature ecosystems, AutoGen and LangChain face the highest maintenance complexity}}. Major version upgrades for AutoGen (\ie from 0.2 to 0.4) are challenging due to changes in the tool definition format. \textbf{\textit{LangChain frequently introduces breaking API changes during upgrades, requiring extensive sample updates, and dependency conflicts (\ie with pydantic) increase its fragility}}. Despite its large community, coordination remains difficult, and consistent code review standards are lacking. Semantic Kernel and LlamaIndex have low upgrade risks, but they can suffer from issues such as incomplete documentation, outdated examples, inconsistent cross-language support (Semantic Kernel), lack of migration guides, and vector storage migration challenges (LlamaIndex). Rapidly iterating frameworks, such as MetaGPT, BabyAGI, and Swarm, face serious stability and compatibility issues. MetaGPT's architecture is fragmented, tool interfaces frequently fail, and maintenance costs rise. BabyAGI's core frequently changes, its plugin system is fragile, and it lacks unified coding standards, making long-term development difficult. Swarm's unstable module dependencies and fragile toolchain further exacerbate maintenance challenges. Camel's model adaptation layer frequently changes, making multimodal expansion complex and costly. LangGraph's tight coupling with LangChain limits independent control, increases the maintenance cost of asynchronous workflows, and lacks independent collaboration standards.

\begin{center}
\begin{tcolorbox}[colback=gray!10,
                  colframe=black,
                  width=\textwidth,
                  arc=1mm, auto outer arc,
                  boxrule=0.5pt,
                 ]
\noindent
\emph{\textbf{Finding 13:} Despite their mature ecosystems, AutoGen and LangChain face the highest maintenance complexity. 
}
\end{tcolorbox}
\end{center}

\begin{figure*}[t]
    \centering
    \includegraphics[width=0.8\linewidth, keepaspectratio]{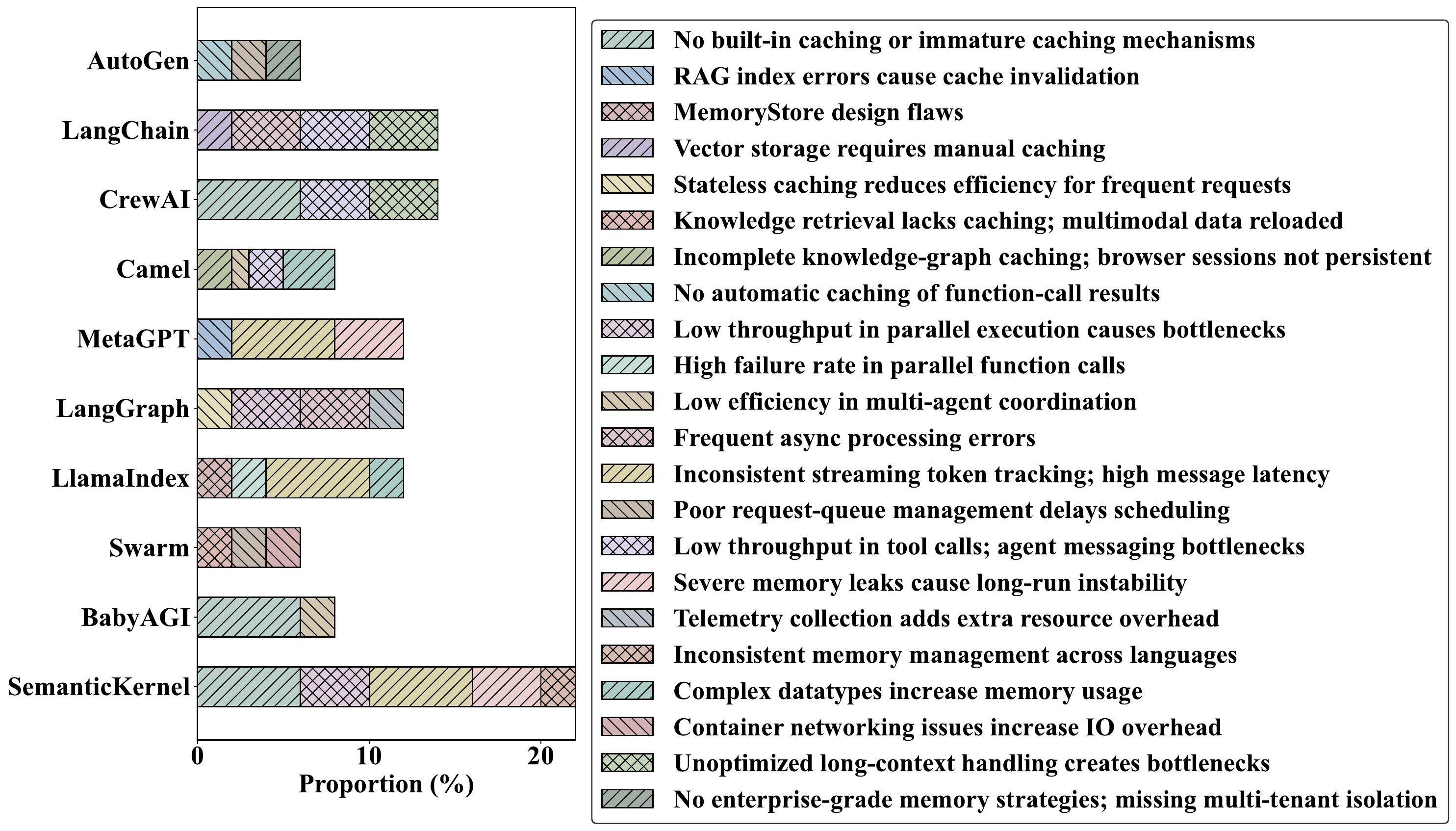}
    \caption{Various Issues Arising in Performance Optimization. SK Denote Semantic Kernel.}
    \label{fig:overview}
    \vspace{-0.1em} %
\end{figure*}

\UseRawInputEncoding
\pdfoutput=1
\section{Discussion}
\label{sec:discussion}

In this section, we analyze developers' discussions about the agent framework, identifying issues related to Model Context Protocol (MCP) and prompting further thinking. We then elaborate on the implications of this research for researchers, practitioners, and educators. Finally, we discuss potential threats to the validity of our research.

\subsection{Model Context Protocol}

From the 8,710 discussions collected during the RQ1 phase, we select and analyze 399 user feedback on MCP tool integration. Overall, MCP has been widely adopted, and its role in engineering workflows is gradually becoming apparent. However, some technical and operational challenges remain.

Model Context Protocol (MCP), introduced by Anthropic in November 2024 ~\cite{Anthropic2024ModelContextProtocol}, is an open standard designed to enable seamless integration between agents and external systems through a client-server architecture. The protocol allows agent assistants to access real-time data and execute external tool operations ~\cite{hou2025model,ehtesham2025survey}. 

On the one hand, adopting MCP can significantly improve engineer productivity. For example, Codename Goose, based on MCP, seamlessly connects APIs and data sources, increasing engineer productivity by approximately 20\%. This effect has been verified in community feedback. Furthermore, the MCP protocol's support for stateful proxies (such as the session persistence feature of the GitHub Issue Manager) simplifies task management and workflow automation, providing convenience for developers.

However, we also identify several negative aspects. First, \textbf{MCP tools fail to operate due to excessively long descriptions}. For example, the description of the sequentialthinking tool exceeds 1,200 characters, causing LibreChat to return a 400 Invalid ``tools.function.description'': string too long error. \textbf{In terms of tool organization and user experience, the current MCP servers do not group tools by server, which makes it harder for users to find and select the right tool when the number of tools is large.} Additionally, MCP configuration and security remain problematic. For instance, the default hardcoded path and insecure credential storage (cline mcp settings.json synchronizes to the cloud) pose risks compared to VSCode's SecretStorage solution. \textbf{Users have widely requested support for improvements in tool organization to reduce redundant system prompts-currently}, prompts reach approximately 11.7k tokens, with MCP instructions accounting for 40\%, leading to LLM processing overload. Finally, \textbf{MCP's scalability in multi-tenant environments is limited}, as each client must start a separate server instance, this issue was also observed in discussions regarding failed integrations with AWS Bedrock and PortKey.

\begin{center}
\begin{tcolorbox}[colback=gray!10,
                  colframe=black,
                  width=\textwidth,
                  arc=1mm, auto outer arc,
                  boxrule=0.5pt,
                 ]
\noindent
\emph{\textbf{Finding 13:}} Excessive prompt overhead, insecure credential storage, and limited multi-tenant scalability remain the main obstacles to MCP adoption.

\end{tcolorbox}
\end{center}

\begin{figure*}[t]
    \centering
    \includegraphics[width=0.9\linewidth, keepaspectratio]{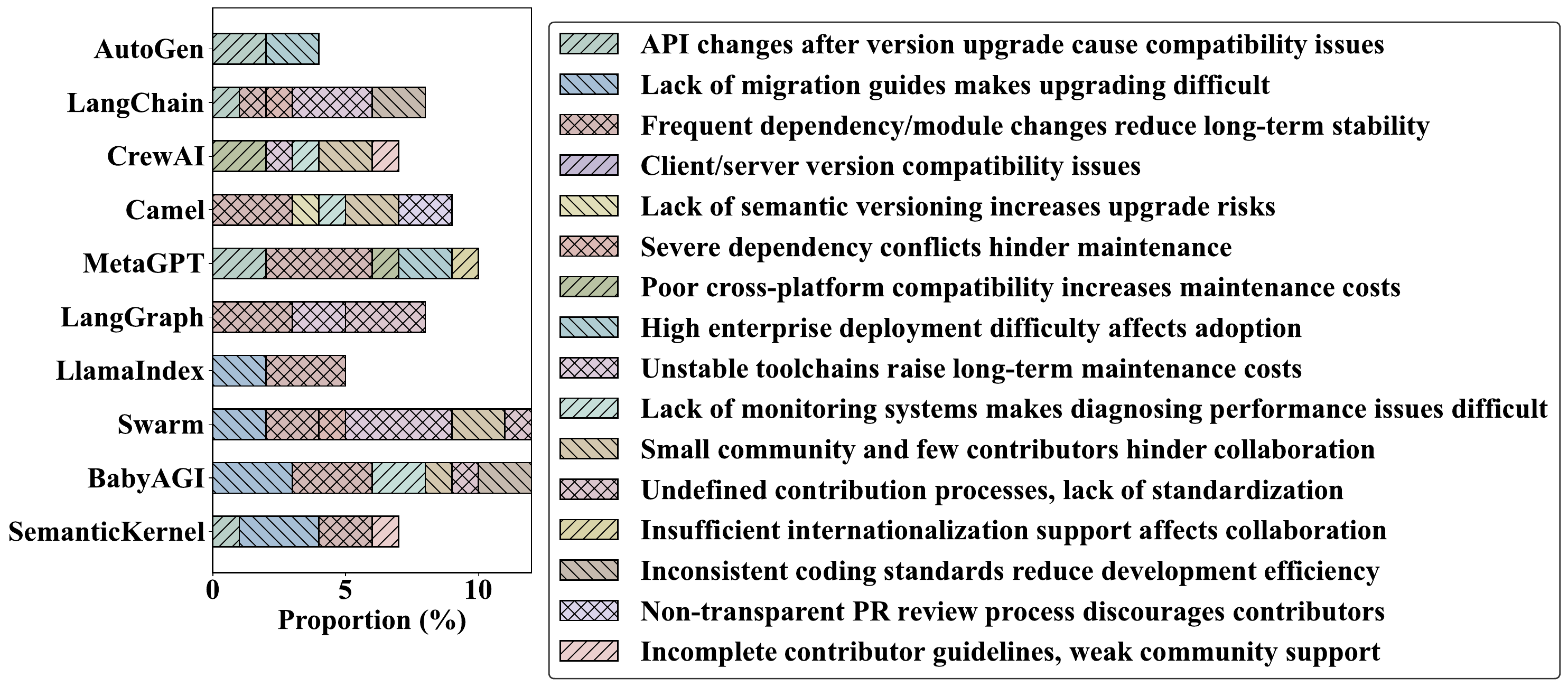}
    \caption{Various Issues Arising From the Maintainability. SK Denote Semantic Kernel.}
    \label{fig:issue5}
    \vspace{0.5em} 
\end{figure*}

\subsection{IMPLICATIONS}
\textbf{For agent developers.}
The rapid proliferation of agent frameworks has created both opportunities and challenges for developers, highlighting the need for informed selection and modular design strategies.
First, developers should not rely solely on GitHub popularity when choosing a framework. Instead, they should also take into account the framework's functional role, its application domain, and the level of community activity and maintenance contributions, in order to mitigate the technical risks of blind adoption. Second, for complex agent systems, developers are advised to adopt modular composition strategies, flexibly combining orchestration frameworks, data retrieval frameworks, and multi-agent collaboration frameworks according to task requirements, thereby achieving complementary strengths. Version dependency conflicts and rapid iterations often lead to community fragmentation, becoming major technical barriers to framework adoption. Therefore, development teams should pay particular attention to version stability and long-term maintenance costs during selection, in order to mitigate risks in future migration and maintenance. Finally, overly abstract frameworks or those lacking practical examples can significantly increase developer cognitive burden and hinder efficiency. Thus, when selecting frameworks, developers must balance learning costs against functional needs to enable more efficient and sustainable system development.

\textbf{For agent framework designers.}
Current agent frameworks still have room for improvement in cross-framework collaboration, interface standardization, performance optimization, and community ecosystem development. Framework designers can reduce the integration cost of combining multiple frameworks by providing standardized interfaces and modular plugin mechanisms, and improve system robustness in multi-agent interactions by introducing built-in features such as message cooling, context deduplication, and dynamic termination. At the same time, to address performance bottlenecks, frameworks should offer native caching mechanisms, concurrency optimization strategies, and resource management tools to enhance stability and security in large-scale workflows. Regarding version management and community ecosystems, framework designers should ensure long-term maintenance and compatibility, provide migration guides and versioning strategies, and avoid fragmentation caused by rapid iterations. Finally, lowering the learning curve remains a key factor in improving framework adoption. Clear documentation, abundant examples, and extensible modular architectures can reduce entry barriers and further promote community engagement and ecosystem growth.

\subsection{THREATS TO VALIDITY}

\textbf{Internal Validity.}
We rely on GPT-4o to automatically summarize and semantically classify over 10,000 developer discussions, which may introduce model-generated bias or semantic misclassification. For example, descriptions of the same functionality being assigned to incorrect categories. Additionally, since LLMs can produce different outputs for the same input, we standardize the generation process by setting the temperature to 0.1 and using top-1 sampling, following established guidelines ~\cite{achiam2023gpt, Guo2024DeepSeekCoderWT}. Moreover, topic tags and dependency files may not fully reflect actual framework usage, particularly for implicit dependencies or customized packages. To mitigate this limitation, we perform multi-source validation by combining framework documentation, example code, and developer discussions, and conduct manual spot checks on key frameworks to capture hidden dependencies and customizations, thereby improving data reliability.

\textbf{External Validity.}
Our sample covers only GitHub open-source projects, excluding closed-source projects, which may limit the generalizability of our conclusions. In enterprise or closed-source environments, framework preferences, dependency management, and collaboration patterns may differ from the open-source community. To mitigate this limitation, we included real-world case studies from industry collaborations to illustrate how these frameworks are applied in practical production environments. In addition, we cross-checked the identified development challenges with public benchmark reports and engineering discussions to ensure that the research findings were not accidental results from a single data source.

\UseRawInputEncoding
\pdfoutput=1

\section{Related Work}
\label{sec:related}
\subsection{LLM-Based Agent}
In recent years, there have been many related works on LLM-Based Agent. LLM-based agents are autonomous systems powered by large language models that perceive, reason, and act in diverse environments ~\cite{kapoor2024ai}. The concept of the agent can be traced back to Denis Diderot's clever parrot theory ~\cite{wang2024survey} , but LLM-based agents have emerged with the breakthrough of the Transformer architecture ~\cite{han2022survey,dou2025alleviating} and the scaling of large language models such as GPT-4 ~\cite{achiam2023gpt}. Unlike early agents ~\cite{jennings1998roadmap} , LLM-based agents ~\cite{xi2025rise} leverage large-scale pretrained language models to support reasoning and decision-making, removing the dependency on manually crafted rules or explicit goal planning. This enables improved generality and contextual understanding across diverse and unstructured tasks. Building on this characteristic, agent systems are increasingly explored in various domains, such as software engineering ~\cite{he2025llm,wang2024agents,xia2024agentless}, drug discoveries ~\cite{zhou2024drug,bou2024acegen}, scientific simulations ~\cite{ghafarollahi2025sciagents,schmidgall2025agent,hong2024data}, and recently general-purpose agent ~\cite{zhang2025agentorchestra,li2024unsupervised,crispino2023agent}.

\subsection{LLM-Based Agent Frameworks } \label{sec:bkg_vul}
Due to the rise of agent systems, many scholars conduct in-depth research on agent frameworks in recent years. Wang et al.~\cite{wang2024survey} propose a unified construction framework that spans key components such as planning, memory, and action modules, also reviewing applications across various domains. Yehudai et al.~\cite{yehudai2025survey} further focus on evaluation strategies for LLM agents, diagnosing gaps in scalability and safety.
Cemri et al.~\cite{cemri2025multiagentllmsystemsfail} present a detailed failure taxonomy for multi-agent LLM systems, identifying 14 failure modes across categories like specification and inter-agent misalignment supported by human annotation. Complementing this, Zhang et al.~\cite{zhang2025agentsafetybenchevaluatingsafetyllm} introduce Agent-SafetyBench, measuring the safety of LLM agents in numerous environments and finding that no agent surpassed a 60\% safety score. Although these studies analyze failure patterns and safety risks, they do not ground their insights in developer-perspective feedback or map failures throughout the lifecycle, an aspect central to our contribution. 
In summary, prior literature has focused on architectural constructs~\cite{shi2024aegis,masterman2024landscape}, safety evaluation~\cite{meyer2024llm,yuan2024r,wang2025comprehensive}, and code-level ~\cite{tang2024codeagent,chen2024coder,takerngsaksiri2024human,10759072}defects in LLM agent systems. Our work complements and extends these by providing a developer-centric empirical analysis that includesss adoption, failure taxonomy, ancomparison of lifecycle-oriented frameworks.

\UseRawInputEncoding
\pdfoutput=1
\section{Conclusion}
\label{sec:conclusion}
This study provides the first large-scale empirical analysis of how developers engage with and adapt AI agent frameworks to develop agent throughout the software development lifecycle (SDLC). By examining ten representative frameworks and analyzing data from thousands of real-world repositories and community discussions, we reveal a comprehensive picture of both their strengths and persistent challenges in practical adoption.
Our findings show that 
the ten LLM-based agent frameworks serve functional roles in four categories, namely basic
orchestration, multi-agent collaboration, data processing, and experimental exploration, and are applied across
ten domains including software development. 
And 96\% of top-starred projects adopt multiple frameworks, highlighting that a single framework can no longer meet the complex needs of agent systems.
We further identify two widely adopted collaboration patterns among agent frameworks. what's more, we propose a taxonomy of agent development challenges in the software development lifecycle (SDLC), covering four domains and nine categories. Finally, we construct a five-dimensional evaluation framework based on agent developer needs to compare the performance of the ten frameworks.

The significance of these findings is that they offer actionable insights for various agent stakeholders, including agent developers and framework designers, by outlining recommendations for framework selection and future enhancement directions.
Our work lays the foundation for understanding agent development based on agent frameworks. By analyzing developer discussions, identifying developer needs, and examining the challenges of building agents using these frameworks, we point out promising directions for future research on agent frameworks.

\section*{Acknowledgements}
This work is supported by the Open Research Fund of The State Key Laboratory of Blockchain and Data Security, Zhejiang University and the National Natural Science Foundation of China (62332004, 62276279) and Guangdong Basic and Applied Basic Research Foundation (2024B1515020032).

\balance
	\bibliographystyle{ACM-Reference-Format}
	\bibliography{Agent-Framework/main}
\end{document}